\documentclass[review,  3p, times]{elsarticle}

\usepackage{outlines}
\usepackage{lineno,hyperref}
\usepackage{tabu}
\usepackage{graphicx}
\usepackage{algpseudocode}
\usepackage{balance}
\usepackage{times}
\usepackage{subfigure}
\usepackage{booktabs}
\usepackage{mathtools}
\usepackage{amsmath}
\usepackage[export]{adjustbox}
\usepackage{amsmath}
\usepackage{mathabx}
\newcommand{\nop}[1]{}
\newtheorem{definition}{Definition}

\newtheorem{lemma}{Lemma}

\usepackage{caption}

\modulolinenumbers[5]
\usepackage[none]{hyphenat}

\usepackage{appendix}

\usepackage{balance} 
\usepackage{graphicx}
\usepackage{balance}  

\usepackage{breqn}
\usepackage{times}
\usepackage{caption}
\usepackage[noabbrev]{cleveref}
\usepackage{multirow}

\usepackage{hyperref}

\usepackage[normalem]{ulem}

\usepackage{subfigure}

\usepackage{tikz}
\newcommand*\circled[1]{\tikz[baseline=(char.base)]{
            \node[shape=circle,draw,inner sep=0.2pt] (char) {#1};}}

\usepackage{color}
\definecolor{Xiang}{rgb}{1,0,0}
\definecolor{Ahmed}{rgb}{0,0,1}

\usepackage[procnumbered,ruled, linesnumbered, vlined]{algorithm2e}
\SetAlFnt{\small}
\journal{Information Systems}

\def\@fnsymbol#1{\ensuremath{\ifcase#1\or \dagger\or \ddagger\or
   \mathsection\or \mathparagraph\or \|\or **\or \dagger\dagger
   \or \ddagger\ddagger \else\@ctrerr\fi}}

\makeatother
    
\providecommand{\keywords}[1]
{
  \textbf{\textit{Keywords---}} #1
}

\begin{document}

\title{Efficient Path Routing Over Road Networks in the Presence of Ad-Hoc Obstacles}

\author{Ahmed Al-Baghdadi$^{\dagger \ddagger}$,
        Xiang Lian$^\dagger$,
        and En Cheng$^{\mathsection}$\\
\textit{$^{\dagger}$Department of Computer Science, Kent State University, Kent, Ohio 44242, USA\\
$^{\ddagger}$Minister of Education in Al-Qadisiyah, Al-Diwaniyah 58001, Iraq\\
$^{\mathsection}$Department of Computer Science, the University of Akron, Akron, Ohio 44325, USA\\}
\emph{\texttt{\{aalbaghd, xlian\}@kent.edu}}\\ \emph{\texttt{\{echeng\}@uakron.edu} }    
}

\begin{abstract}
Nowadays, the path routing over road networks has become increasingly important, yet challenging, in many real-world applications such as location-based services (LBS), logistics and supply chain management, transportation systems, map utilities, and so on. While many prior works aimed to find a path between a source and a destination with the smallest traveling distance/time, they do not take into account the quality constraints (e.g., obstacles) of the returned paths, such as uneven roads, roads under construction, and weather conditions on roads. Inspired by this, in this paper, we consider two types of ad-hoc obstacles, keyword-based and weather-based obstacles, on road networks, which can be used for modeling roads that the returned paths should not pass through. In the presence of such ad-hoc obstacles on roads, we formulate a \textit{path routing query over road networks with ad-hoc obstacles} (PRAO), which retrieves paths from source to destination on road networks that do not pass ad-hoc keyword and weather obstacles and have the smallest traveling time. In order to efficiently answer PRAO queries, we design effective pruning methods and indexing mechanism to facilitate efficient PRAO query answering. Extensive experiments have demonstrated the efficiency and effectiveness of our approaches over real/synthetic data sets.
\end{abstract}

\maketitle


\renewcommand{\baselinestretch}{1.5} 
\section{Introduction}
\begin{figure}[t!]
\vspace{3ex}%
  \centering
  \scalebox{0.4}[0.4]{\includegraphics{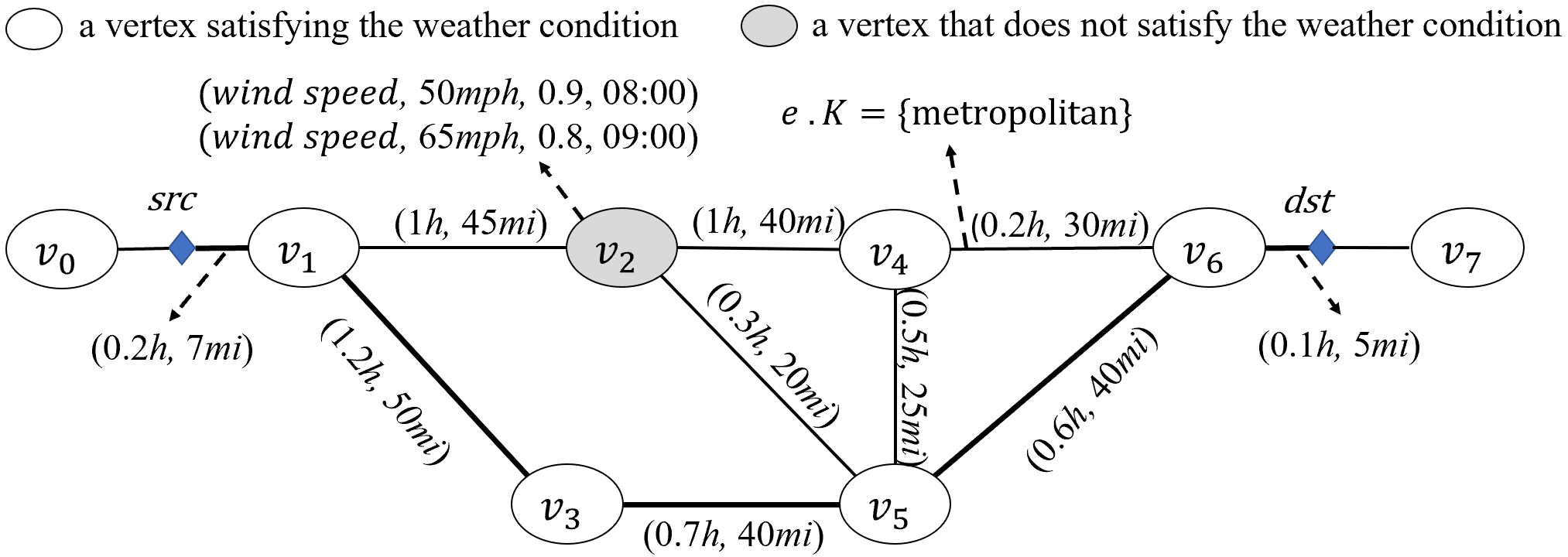}}
     \caption{An example of a road network, illustrating wind speeds of vertex $v_2$ at 08:00 am and 09:00 am, and edge $e$ $(=v_4v_6)$ passing through a metropolitan area.}
     \label{fig:EX}
\end{figure}

In many real-world applications \cite{Dijkstra59,Kung86,agrawal1990direct,Ioannidis93,huang1997integrated,Li05b,Ding08,hua2010probabilistic} such as location-based
services (LBS), supply chain management, map services,
transportation systems, and so on, one classical and important
problem is to find the best path (e.g., with the smallest traveling
time or distance) between a source and a destination over road
networks. Figure \ref{fig:EX} shows an example of a road network,
where line segments (e.g., $v_1v_2$) represent roads, and nodes
(e.g., $v_1$) denote intersection points of roads. In this example,
given two points, $src$ and $dst$, on road networks, the path
routing task is to find a path from source $src$ to destination $dst$ with the smallest
traveling time.

Different from prior works on the path routing problem, in this
paper, we will take into account two specific types of obstacles on
road networks, keyword-based and weather-based obstacles, and
retrieve a path with the smallest traveling time that does not pass
through some user-specified ad-hoc obstacles (or road segments).


Below, we give an example of path routing in the real applications
of transferring chemical supplies.

\setlength{\textfloatsep}{0pt}

\begin{table}[t!]
\centering\footnotesize
\begin{tabular}{||c|c|c||c|c||}\hline
    {\bf V. ID} & {\bf Wind Speed at 8am} & {\bf Conf.} & {\bf Wind Speed at 9am} & {\bf Conf.}\\ \hline\hline
    $v_0$ & 10 mph & 0.9 & 15 mph& 1\\\hline
    $v_1$ & 15 mph & 0.8 & 20 mph & 0.9\\\hline
    \uwave{{\bf $v_2$}} & \uwave{{\bf 50 mph}} & \uwave{{\bf 0.9}} & \uwave{{\bf 65 mph}} & \uwave{{\bf 0.8}}\\\hline
    $v_3$ & 20 mph & 0.9 & 25 mph & 0.8\\ \hline
    $v_4$ & 35 mph & 0.9 & 40 mph & 0.9\\\hline
    $v_5$ & 20 mph & 1 & 25 mph & 0.9\\\hline
    $v_6$ & 10 mph & 0.8 & 15 mph & 1\\\hline
    $v_7$ & 10 mph & 0.9 & 10 mph & 1\\\hline
\end{tabular}
\caption{Weather forecasting in the example of Figure \ref{fig:EX}.}
\label{table:weather}\vspace{3ex}
\end{table}


\vspace{1ex}\noindent {\bf Example 1. (Path Routing for Chemical Transportation)}. {\it As illustrated in Figure \ref{fig:EX}, assume that a company wants to deliver dangerous chemical compounds from a source, $src$, to a destination, $dst$, on road networks. Due to the potential spread of hazardous chemicals, this company would like to schedule an optimal path to transfer chemical supplies, which satisfies the following conditions: (1) the path should not pass by those roads close to areas with high population densities; (2) the wind speed on the path should not exceed 40mph; and (3) the retrieved path has the smallest traveling time.

In this example, the company can online specify obstacle keywords such
as ``metropolitan'' (i.e., ad-hoc keyword-based obstacles) and
require wind speeds on roads smaller than or equal to 40mph (i.e.,
ad-hoc weather-based obstacles). Given road networks in Figure \ref{fig:EX} and weather conditions in Table \ref{table:weather}, we can obtain a path that
does not visit roads with obstacle keyword ``metropolitan'', with
wind speeds always not exceeding 40mph of high confidences, and
having the smallest traveling time. In this example, one candidate
path can be $src \rightarrow v_1 \rightarrow v_3 \rightarrow v_5 \rightarrow v_6 \rightarrow
dst$.$\hfill\blacksquare$\\

}

In the example above, keyword-based obstacles indicate properties of roads, such as ``uneven roads'', ``roads under construction'',
``roads close to downtown areas'', ``roads with falling rocks'', ``roads with deer'', and so on. In Figure \ref{fig:EX}, edge $e$ ($=v_4v_6$) has a keyword ``metropolitan''. On the other hand, weather-based obstacles are highly related to the predicted weather conditions in the future, such as temperature, humidity, or wind speeds (as depicted in Table \ref{table:weather}). Moreover, due to the accuracy of weather forecasting, each predicted weather value is often associated with a confidence. In Table
\ref{table:weather}, the wind speed at vertex $v_2$ is 50mph at 8am with probability 0.9, and 65mph at 9am with probability 0.8.

\vspace{1ex}\noindent {\bf Example 2. (Supply Chain Management)}. {\it In a recent study by the Atmospheric and Environmental Research Center \cite{azcentral}, thirty percent (30\%) of the U.S. gross domestic product is affected by the weather condition. One of the main weather effects comes from delivery issues. According to the article, weather conditions can make it hazardous and even impossible for delivery trucks to transport products to distributors and retailers. Weather conditions such as snow, ice, high temperature, high wind speed, and heavy rain can slow and stop transportation, making products such as groceries unavailable when they are most needed. 

Therefore, in the real application of supply chain management, it is very important for many users (such as logistics companies or delivery drivers) to find an optimal path between source and destination for the product delivery, by considering weather conditions and road status. For example, a user can specify constraints (i.e., weather-based obstacle) with respect to temperature (e.g., temperature $< 45^{\circ}$), wind speed (e.g., wind speed $< 40mph$), and/or any other weather-related obstacles at the query time, such that the retrieved path will avoid passing through roads that do not satisfy the specified weather conditions. Furthermore, the user can specify some keywords to be avoided in the retrieved path (i.e., keyword-based obstacles). For instance, the truck driver can specify ``city road'' as the obstacle keyword(s) to avoid roads passing through cities in the path finding process. $\hfill\blacksquare$\\

}

Note that, the aforementioned keyword-based and weather-based
obstacles are ad-hocly given, in the sense that different path
routing problems may specify different obstacle criteria, including
an arbitrary set of user-specified obstacle keywords and
domain-specific weather constraints for the paths to be returned.
Thus, in the presence of such ad-hoc obstacles, it is very challenging to effectively and efficiently compute the path over large-scale road networks.


Inspired by the motivation examples above, in this paper, we formalize a \textit{path routing query with ad-hoc obstacles} (PRAO), which
retrieves a path that does not pass through ad-hoc keyword-based and weather-based obstacles and
achieves the smallest traveling time. As mentioned above, the PRAO problem has many potential users (e.g., logistics companies, delivery drivers, etc.) and high/frequent demands (e.g., map utility and services, mobile APP requests by drivers or logistics managers).



{Due to the large amount of weather/keyword information over large-scale road networks (exponential number of possible worlds for the predicted weather data), it is rather challenging to efficiently process PRAO queries. One straightforward method to answer PRAO queries is to enumerate all possible paths between $src$ and $dst$ (e.g., via $A^*$ algorithm \cite{Dijkstra59}), check keyword-based and weather-based obstacle constraints on each road segment of paths, and return the one with the smallest traveling time. However, this straightforward method is not time-efficient, due to a large number of possible paths and the costly computations of checking weather predictions for any point on paths. Therefore, it cannot efficiently handle frequent query requests from users, and we need to design specific techniques (e.g., pruning/indexing/query processing) to tackle the PRAO problem, which are the main contributions of our work.
}

\nop{
{\color{Xiang} {\bf (Ahmed, this should not be in the introduction, this should be in experimental section!!!!!)
To confirm the efficiency of our proposed PRAO approach, we compared our PRAO approach with two straightforward baseline algorithms. The first baseline algorithm, namely, $A^*$, that is a variant of Dijkstra's $A^*$ algorithm, where we adopt it to handle edges with ad-hoc obstacles.
The second baseline, namely $filterfirst$, which first online filters out all edges that contain ad-hoc obstacles, and then conducts the $A^*$ algorithm over the remaining road networks (excluding those with obstacles) ) to find the shortest path between the source and the destination. As shown in Section \ref{sec:exper}, our PRAO approach can outperform the two baselines by orders of magnitude. }
}
}

To tackle the PRAO problem, in this paper, we propose three effective pruning methods, keyword-based,
weather-based, and traveling time pruning, to filter out false
alarms of candidate paths and reduce the PRAO search
space. We can ensure that our proposed method does not miss any optimal route nor produces false dismissals (i.e., we always prune those false alarms of paths containing ad-hoc keyword-based obstacles and/or weather-based obstacles with high confidences, and retrieve actual PRAO answers satisfying PRAO query predicates). Furthermore, we design a variant of R$^*$-tree index over road networks with obstacles, which supports dynamic updates of weather information. That is, we index road-network edges in a variant of R$^*$-tree where each tree node represents a group of spatially closed edges (roads), and all the information about weather predictions and keywords is summarized and stored in each node. Finally, we propose an efficient algorithm to facilitate PRAO query answering via the index. Our proposed pruning methods w.r.t. index nodes can guarantee no false dismissals.

\nop{ 
{\color{Xiang} \bf (Ahmed, the weather should be integrated into weather section without redundancy!)
In this work, we collect real weather information at intersection points of road networks from Dark Sky \cite{dark}, as we discuss in Section \ref{sec:prob_def}. Then, we interpolate/estimate weather conditions at unknown areas from known weather at other areas. 
For a moving vehicle on the edge $e_i$ $(= v_jv_k)$ departing from $v_j$ at a timestamp $t_{dep}$ toward $v_k$ (with an expected arrival time $t_{arr}$), we can interpolate/estimate weather conditions on any points $o_l \in e_i$ at any timestamp 
$t$ $(\in [t_{dep}, t_{arr}])$, according to the weather information known at $v_j$ and $v_k$, as described in Section \ref{sec:weather_est}.
}
}

In this paper, we make the following contributions.

\begin{enumerate}
\item We formulate the PRAO problem over road networks with ad-hoc obstacles in Section
\ref{sec:prob_def}, and present the weather estimation for any location on roads in Section \ref{sec:weather_est}.

\item We present three effective pruning strategies to reduce the PRAO search space in Section
\ref{sec:pruning}.

\item We propose effective indexing and efficient PRAO query answering algorithms in Sections  \ref{sec:query_processing} and \ref{sec:QPalgorithm}, respectively.


\item We demonstrate through extensive experiments the efficiency and effectiveness of our
proposed PRAO query answering approach in Section
\ref{sec:exper}.

\end{enumerate}

Section \ref{sec:related} discusses related works on queries over road networks, probabilistic data management, and queries in the presence of obstacles. Section \ref{sec:conclusion} concludes the paper.

\section{Problem Definition}
\label{sec:prob_def}

\subsection{Data Model for Road Networks}

First, we formally present the data model for road networks as follows.

\begin{definition} \textbf{(Road Networks)} A road network is represented by a graph $G=(V,E,\phi)$, where $V$ is a set of vertices $v_1,v_2,\ldots,$ and $v_{|V|}$, $E$ is a set of edges $e_1, e_2,\ldots,$ and $e_{|E|}$, and $\phi$ is a mapping function: $V \times V \rightarrow E$, where each edge $e_i$ ($=v_jv_k$) is associated with a traveling time $e_i.w$ and a traveling distance $e_i.len$ ($=dist(v_j, v_k)$).\label{def:road_network}
\end{definition}

In Definition \ref{def:road_network}, edges $e_i$ (for $1\leq i \leq |E|$) in a road network $G$ represent roads, whereas vertices $v_j$ (for $1\leq j \leq |V|$) correspond to road intersection points. Figure \ref{fig:EX} shows an example of road networks, where vertices $v_0\sim v_7$ represent intersection points on road networks. Edge $e=v_4v_6$ is a road segment connecting two intersection points $v_4$ and $v_6$, associated with: (1) the traveling time $e.w=0.2$ $hour$, and (2) the edge length, $e.len = dist(v_4,v_6)=30$ $miles$.  

For simplicity, in this paper, we consider a \textit{static} traveling time $e_i.w$ for each road segment $e_i$. For time-varying traveling times on roads, the data model is more complex, and the problem over such road networks is more challenging and worthy of investigation. We would like to leave the interesting topic of considering our problem over  road networks with dynamic traveling times on roads as our future work.


\subsection{Data Model for Ad-Hoc Obstacles}
\label{subsec:obstacle_model}

In this subsection, we provide formal definitions of two types of ad-hoc obstacles on road networks, ad-hoc keyword-based and weather-based obstacles. 

\noindent {\bf Ad-Hoc Keyword-based Obstacles:} We define ad-hoc keyword-based obstacles below.

\begin{definition} \textbf{(Ad-Hoc Keyword-based Obstacles)}
On road networks $G$, each edge $e_i$ is associated with a set, $e_i.K$, of obstacle keywords. Given a set, $S$, of user-specified keywords, an edge $e_i$ on road network $G$ is said to be an \textit{ad-hoc keyword-based obstacle}, if $e_i.K$ contains some obstacle keywords 
in $S$ (i.e., $e_i.K \bigcap S \neq \emptyset$).
\end{definition}

In particular, ad-hoc keyword-based obstacles can be on roads (edges) that pass through specific sites (e.g., metropolitan areas with high population densities), or on roads that travelers prefer not to pass through (e.g., uneven roads).

In Figure \ref{fig:EX}, the edge $v_4v_6$ (i.e., $e$) has a keyword set  \{metropolitan\}. If a user-specified set, $S$, of obstacle keywords is \{uneven, metropolitan\}, then edge $v_4v_6$ contains an ad-hoc keyword-based obstacle.

\noindent {\bf Ad-Hoc Weather-based Obstacles:} Next, we give the definition of ad-hoc weather-based obstacles.

\begin{definition} \textbf{(Weather Predictions)} Weather predictions for a point $o_l \in e_i$ are given by a set, $W(o_l)$, of quadruples, in the form $(W\_type, W\_val(o_l, t_l), p(o_l,t_l), t_l)$, where $W\_type$ represents the type of weather conditions, $W\_val(o_l,t_l)$ shows the possible value of the weather condition at the timestamp $t_l$, and $p(o_l,t_l)$ denotes the confidence of weather value $W\_val(o_l,t_l)$.\label{def:weather}
\end{definition}
\sloppy
 In Definition \ref{def:weather},  the type, $W\_type$, of weather conditions can be wind speed, temperature, and so on. As an example in Figure \ref{fig:EX}, the quadruple, $(wind\text{ }speed, 50mph, 0.9, 8\text{:}00)$, at vertex $v_2$ indicates that the predicted wind speed (i.e., $W\_type$) 50mph at 8am is accurate, with prediction confidence 0.9.

 Note that, here we can obtain hourly weather prediction data from data sources such as Dark Sky \cite{dark}. For instance, highway I24 might have a 70\% probability of wind speed 50mph at timestamp 5am. Moreover, weather data are dynamically updated on an hourly basis. That is, the newly predicted weather data arrive every hour, whereas expired weather data can be removed from the system.


Furthermore, it is not feasible to predict the weather information at every point $o_l$ of road-network edges $e_i$, thus, in this work, we assume that we can obtain the predicted weather information at the two endpoints $v_j$ and $v_k$ of each road-network edge $e_i$ $(= v_jv_k)$. We can estimate weather conditions (as well as their confidences) at any point $o_l$ on the edge $e_i$ by using weather interpolation techniques.
In particular, for a moving vehicle on the edge $e_i$ $(=v_jv_k)$ departing from $v_j$ at a timestamp $t_{dep}$ toward $v_k$ (with an expected  arrival  time $t_{arr}$), we  can  interpolate/estimate  weather  conditions  on  any  points $o_l\in e_i$ at  any timestamp $t$ ($\in [t_{dep},t_{arr}]$), according to the weather information known at $v_j$ and $v_k$, which will be discussed later in Section \ref{sec:weather_est}.

\begin{definition} \textbf{(Ad-Hoc Weather-based Obstacles)} Given weather predictions $W(o_l)$ on road network $G$, an ad-hoc weather threshold $\epsilon$, and a probabilistic threshold $\alpha$, any point $o_l$ on edge $e_i$ is said to be an \textit{ad-hoc weather-based obstacle} at a timestamp $t_l$, if it holds that $Pr\{W\_val(o_l, t_l) > \epsilon\}\geq \alpha$.\label{def:weather_obstacle}
\end{definition}

Intuitively, a point $o_l$ on an edge $e_i$ is an ad-hoc weather-based obstacle, if and only if the value of the weather condition (e.g., a wind speed 50mph or a temperature 50$^\circ$C) at $o_l$ is greater than a threshold $\epsilon$ with high confidence (i.e., $Pr\{W\_val(o_l, t_l) > \epsilon\}\geq \alpha$). Due to bad weather at $o_l$ or the required weather conditions on the traveling path, those edges $e_i$ containing weather-based obstacle $o_l$ will be prohibited to be passed through. 


\nop{

\begin{definition} \textbf{(Safe Road Trip, SRT)} Given a trip on a road $e_i$ $(=v_jv_k)$ that departs from $v_j$ at a timestamp $t_{dep}$ and arrives at $v_k$ at a timestamp $t_{arr}$, if any point $o_l$ during the trip on road segment $e_i$ does not contain ad-hoc weather-based obstacles nor the edge $e_i$ contains ad-hoc keyword-based obstacles, we say that this trip is safe, denoted as SRT($e_i,[t_{dep},t_{arr}]$), where $t_{arr}= t_{dep} + e_i.w$.\label{def:SRT}
\end{definition}

In Definition \ref{def:SRT}, SRT($e_i,[t_{dep},t_{arr}]$) is true, if for any point $o_l \in e_i$ during the trip, it holds that $Pr\{W\_val(o_l,t_l)\leq \epsilon\}\geq \alpha$ and $e_i.K\bigcap S=\emptyset$, where $S$ is the set of obstacle keywords.

}


\subsection{The PRAO Problem Definition}
In this subsection, we define the path routing over road networks in the presence of ad-hoc keyword-based and weather-based obstacles (i.e., the PRAO problem).
\sloppy
\begin{definition} \textbf{(Path Routing Over Road Networks with Ad-Hoc Obstacles, PRAO)} Assume that we have a road network $G$
with weather conditions $W(v_j)=\{(W\_type, W\_val(v_j, t_j), p(v_j, t_j), t_j)\}$ at any vertex $v_j \in V$ in $G$ at future timestamps $t_j$,
a query weather type $W\_type$,
a weather threshold $\epsilon$,
a probabilistic threshold $\alpha$,
a set, $S$, of obstacle keywords,
a source vertex $src$,
and a destination vertex $dst$.
A \textit{path routing query with ad-hoc obstacles} (PRAO) finds the best path, $Path$, from $src$ to $dst$ that satisfies the following three criteria:
\begin{itemize}
\item for any point $o_l$ on path, $Path$, with arrival time $t_l$, it holds that:
\begin{eqnarray}
Pr\left \{  W\_val(o_l,t_l)> \epsilon \right \} < \alpha;\label{eq:eq0}
\end{eqnarray}
\item for any obstacle keyword $s\in S$ and edge $e_i\in Path$, it holds that $e_i.K$ does not contain $s$ (i.e., $S \bigcap e_i.K=\emptyset$), and;
\item the total traveling time of path, $Path$, that is, $\sum_{\forall e_i \in Path} e_i.w$, is minimized. 
\end{itemize}\label{def6}
\end{definition}

In Definition \ref{def6}, the PRAO problem searches for a path, $Path$, from $src$ to $dst$ that satisfies good weather conditions with high probabilities, avoids passing through edges with keyword obstacles, and has the minimum traveling time.

\begin{table}[t!]
\centering{\footnotesize
\begin{tabular}{||p{2cm}||p{10.5 cm}||}
     \hline
    {\bf Symbol} & \qquad\qquad\qquad\qquad\qquad\qquad {\bf Description} \\
     \hline
     \hline
     $G$& a road network (graph)\\
     \hline
     $V$& a set of vertices\\
     \hline
     $E$& a set of edges\\
     \hline
     $e_i.w$ & the traveling time of the edge $e_i$\\
     \hline
     $e_i.len$ & the length of road segment $e_i$\\     
     \hline
     $o_l$& an object or an ad-hoc obstacle\\
     \hline
     $dist(v_j,v_k)$ & the Euclidean distance between $v_j$ and $v_k$\\
     \hline
     $dist_N(v_j,v_k)$ & the shortest path distance between $v_j$ and $v_k$ \\
     \hline
     $e_i.K$ &a set of obstacle keywords on edge $e_i$ \\
     \hline
     $S$ & a set of user-specified obstacle keywords \\
     \hline     
     $W(v_j)$ & the weather predictions at vertex $v_j$\\
     \hline
     $W\_type$ & the type of weather condition\\
     \hline
     $W\_val(v_j, t_j)$& the weather value of vertex $v_j$ at timestamp $t_j$\\
     \hline
     $\epsilon$ &an ad-hoc weather threshold\\ 
     \hline
     $p(v_j,t_j)$ & the confidence of weather condition predictions at the intersection point $v_j$ at timestamp $t_j$\\
     \hline
    \end{tabular}}
\caption{Frequently used notations}
\label{table1}\vspace{3ex}
\end{table}

\subsection{Challenges}

The main challenges of the PRAO problem are threefold. First, in the PRAO problem, there are many edges with keywords, which may potentially be ad-hoc obstacles on road networks (i.e., containing ad-hoc obstacle keywords). It is not trivial how to efficiently compute the best paths online that do not pass through these keyword-based obstacles. Second, the weather conditions are available only at road intersection points (i.e., vertices) with probabilistic confidences, and it is not trivial to accurately estimate weather conditions at any points on roads. Moreover, the weather-based obstacles are ad-hocly specified by users at the query time, and it is rather challenging to efficiently retrieve valid paths by avoiding ad-hoc weather-based obstacles. Third, road networks are usually of large scale, and  there are a large number of possible paths between $src$ and $dst$, which are not time-efficient to enumerate. 

Inspired by the challenges above, in this paper, we will propose an efficient query processing approach to answer PRAO queries. The proposed approach follows a pruning procedure to filter out false alarms of paths and reduce PRAO search space. We will also design an efficient query answering algorithm to answer PRAO queries over road networks.

Table \ref{table1} depicts the commonly used symbols and their descriptions in this paper.

\section{Weather Condition Estimation}
\label{sec:weather_est}

 As mentioned in Section \ref{subsec:obstacle_model}, for each edge $e_i$ $(=v_jv_k)$, we assume that the predicted weather conditions with confidences are available at vertices $v_j$ and $v_k$. However, the weather conditions at other points on edge $e_i$ (excluding $v_j$ and $v_k$) are unknown. Therefore, in this section, we will discuss how to estimate the unknown weather condition (and its confidence as well) for any point on edges in road networks, based on known weather conditions at vertices.

\nop{
In this section, we will discuss how to estimate the weather condition (and its confidence as well) for any point on edges in road networks. Specifically, we consider
a moving vehicle on an edge $e_i$ $(=v_jv_k)$ departing from $v_j$ at timestamp $t_{dep}$ toward $v_k$ with velocity $\frac{e_i.len}{e_i.w}$ $(= \frac{dist(v_j,v_k)}{e_i.w})$. For timestamp $t_l$ (where $t_{dep}\leq t_l\leq t_{arr}$), we will estimate weather conditions at $o_l\in e_i$ according to the known weather conditions at vertices $v_j$ and $v_k$. 
}


Without loss of generality, consider a moving vehicle on an edge $e_i$ $(=v_jv_k)$ departing from $v_j$ at timestamp $t_{dep}$ toward $v_k$ with velocity $\frac{e_i.len}{e_i.w}$ $(= \frac{dist(v_j,v_k)}{e_i.w})$. At timestamp $t_l$ ($\in [t_{dep}, t_{arr}]$), we will estimate weather conditions at $o_l\in e_i$, inferred from the known (predicted) weather conditions, $W\_val(v_j, t_l)$ and $W\_val(v_k, t_l)$, at vertices $v_j$ and $v_k$, respectively.

Since the predicted weather condition $W\_val(v_j, t_l)$ at vertex $v_j$ (or $W\_val(v_k, t_l)$ at vertex $v_k$) is associated with a probability $p(v_j, t_l)$ (or $p(v_k, t_l)$), the predicted weather condition at $v_j$ (or $v_k$) can be either accurate or inaccurate. Following the \textit{possible worlds} semantics \cite{Dalvi07}, we will consider four cases below (i.e., 4 possible worlds, based on whether or not the predicted weather data at $v_j$ and $v_k$ appear in the real world), and estimate the weather condition at any location $o_l \in e_i$ in each of these four cases.

{\bf Case \circled{A}} {\bf(The predicted weather values at both vertices $v_j$ and $v_k$ are accurate)}. In this case, based on accurate weather predictions at vertices $v_j$ and $v_k$, we can estimate/interpolate the weather information at $o_l \in e_i$ by using \textit{Inverse Distance Weighting} (IDW) interpolation \cite{shepard1968two}. The IDW interpolates weather information at unknown points $o_l$ from weighted average of the values available at the known points $v_j$ and $v_k$ as follows: 
\begin{equation}
W\_val(o_l, t_l) =
\frac {dist(o_l,v_k)\cdot W\_val(v_j,t_l)+dist(o_l,v_j)\cdot W\_val(v_k,t_l)}{dist(o_l,v_j)+ dist(o_l,v_k)},\nonumber
\label{eq1} 
\end{equation}

\noindent where $o_l$ is the point at which we are interested in finding its weather, $v_j$ and $v_k$ denote the vertices with known weather data, and $dist(x, y)$ is the Euclidean distance between $x$ and $y$. 

Intuitively, when point $o_l$ becomes closer to $v_j$ (or $v_k$), its estimated weather value $W\_val(o_l, t_l)$ will be more similar to that of $v_j$ (or $v_k$).

The confidence that point $o_l$ has the weather value $W\_val(o_l,t_l)$ can be derived from the confidence that both vertices $v_j$ and $v_k$ have accurate weather predictions. That is, we have:
$$p(o_l,t_l)=p(v_j,t_l) \cdot p(v_k,t_l),$$ where $p(o_l,t_l)$ is the confidence of the value of weather conditions of the point $o_l$ at timestamp $t_l$.

\begin{figure}[t!]
\centering
\subfigure[][{\small weather conditions at vertices $v_j$ and $v_k$}]{                    
\scalebox{0.5}[0.5]{\includegraphics{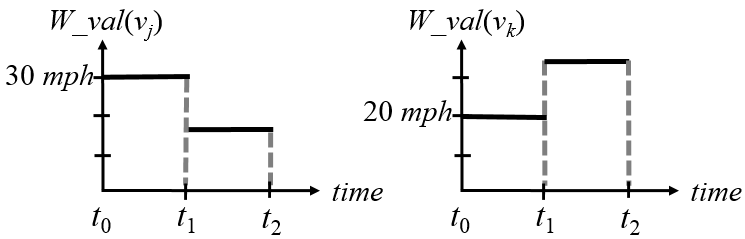}}
}
\subfigure[][{\small the interpolation of weather conditions at $o_l$ from known weather conditions at $v_j$ and $v_k$}]{                    
\scalebox{0.8}[0.8]{\qquad\qquad\qquad\includegraphics{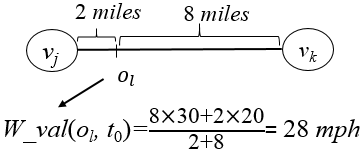}\qquad\qquad\qquad}
}
\caption{An example of the weather interpolation (Case \raisebox{.5pt}{\textcircled{\raisebox{-.9pt} {A}}}).} \label{fig:WS} \vspace{3ex}
\end{figure}

Figure \ref{fig:WS} illustrates an example of the interpolation of weather information at point $o_l \in e_i$, where wind speeds at $v_j$ and $v_k$ are 30 mph and 20 mph, respectively. Based on the distance from point $o_l$ to $v_j$ or $v_k$ (i.e., 2 miles and 8 miles, resp.), we can estimate the wind speed at $o_l$ as $W\_val(o_l, t_l) = \frac{8\times 30 + 2\times 20}{2+8} = 28$ mph.

{\bf CASE \circled{B}} {\bf(The predicted weather condition at vertex $v_j\in e_i$ is inaccurate, and that at vertex $v_k \in e_i$ is accurate)}. In such a case, we rely on known and accurate information at vertex $v_k$ to estimate value of weather condition at $o_l$ on edge $e_i$ as follows:
$$
W\_val(o_l,t_l)= W\_val(v_k,t_l).
$$
The confidence of the estimated weather value at $o_l$ can be derived as follows:
$$
p(o_l,t_l)=(1-p(v_j,t_l)) \cdot p(v_k,t_l).
$$

{\bf CASE \circled{C}} {\bf(The predicted weather condition at vertex $v_j\in e_i$ is accurate, and that at vertex $v_k \in e_i$ is inaccurate)}. We estimate values of weather conditions at $o_l$ by using the known and accurate weather information at $v_j$ as follows:
$$
W\_val(o_l,t_l)= W\_val(v_j,t_l).
$$
Similarly, we derive the confidence of the weather value at $o_l$ as follows:
$$
p(o_l,t_l)=p(v_j,t_l) \cdot (1- p(v_k,t_l)).
$$

{\bf Case \circled{D} (The predicted weather conditions at both $v_j$ and $v_k$ are inaccurate).} In this case, the weather at any point $o_l$ on edge $e_i$ is unknown (i.e., it can be any value), with probability: $$p(o_l, t_l) = (1-p(v_j, t_l))\cdot (1-p(v_k, t_l)).$$

For any point $o_l$ on the edge $e_i$ at timestamp $t_l$, we can obtain different possible weather values, $W\_val(o_l,t_l)$, corresponding to the four cases (\circled{A}, \circled{B}, \circled{C}, and \circled{D}) above, each of which is associated with a confidence value $p(o_l, t_l)$.

\section{Pruning Strategies}
\label{sec:pruning}
\label{section4}
In this section, we present three pruning strategies, \textit{keyword-based}, \textit{weather-based}, and \textit{traveling time pruning}. The pruning strategies utilize the properties of the PRAO problem to improve the efficiency of PRAO query processing.

\subsection{Keyword-based Pruning}


In this subsection, we present an effective \textit{keyword-based pruning} method, which can reduce the PRAO search space, under the constraint of keyword-based obstacles.

Given a set, $S$, of obstacle keywords that should not appear on the path from $src$ to $dst$, an edge, $e_i$, can be safely pruned, if there exists an obstacle keyword $s\in S$ in the keyword set, $e_i.K$, of edge $e_i$ (i.e., $e_i.K\bigcap S \neq \emptyset$). Intuitively, if an edge $e_i$ contains keyword-based obstacles $s \in S$, then we can safely prune this edge. 


\begin{lemma} {\bf (Keyword-based Pruning)}
Given an edge $e_i \in E$ and a set, $S$, of ad-hoc obstacle keywords, if $e_i.K \bigcap S \neq \emptyset$ holds, then edge $e_i$ can be safely pruned.\label{lemma:lem1}
\end{lemma}
\noindent {\bf Proof.} For details, please refer to Appendix A. \qquad $\square$\\

\nop{
\noindent {\bf Proof.} Since $e_i.K \bigcap S \neq \emptyset$ holds, we can infer that edge $e_i$ is associated with some obstacle keywords in $S$. Therefore, edge $e_i$ cannot appear on any path from $src$ to $dst$, based on the PRAO problem definition (given in Definition \ref{def6}). \qquad $\square$
}

In order to enable keyword-based pruning, we hash the keyword set $e_i.K$ of each edge $e_i$ (or the set, $S$, of obstacle keywords) into a bitmap array \cite{chan1998bitmap}, where each bit in the bitmap represents a hashed keyword and the bit value represents the existence of the keyword. If the bit value at a position in the bitmap is equal to 1, then the keyword mapped to that position exists; otherwise (i.e., the bit is 0), the corresponding keyword does not exist.


Therefore, we can test if a keyword $s\in S$ appears on the edge $e_i$, by applying the AND operation between the two bitmaps of $e_i.K$ and $S$. If the AND result (i.e., a bitmap) is nonzero, it indicates that sets $e_i.K$ and $S$ share common obstacle keywords (i.e., $e_i.K\bigcap S \neq \emptyset$), and edge $e_i$ can be safely pruned.


\subsection{Weather-based Pruning}
\label{Weather-basedPruning}
In this subsection, we propose a \textit{weather-based pruning} method to filter out edges, $e_i$, that do not satisfy the constraint of weather conditions in Definition \ref{def6}. The basic idea of our weather-based pruning method is to prune those edges, $e_i$, with bad weather (or alternatively, with
low chances of having good weather).



\begin{lemma} {\bf (Weather-based Pruning)} Given a point $o_l$ on edge $e_i$, at timestamp $t_l$, we denote $UB\_Pr\{W\_val(o_l,t_l)\leq \epsilon\}$ as an upper bound of probability $Pr\{W\_val(o_l,t_l)\leq\epsilon\}$. If it holds that $UB\_Pr\{W\_val(o_l,t_l)\leq\epsilon\}< 1-\alpha$, then edge $e_i$ can be safely pruned.\label{lemma:lem2}
\end{lemma} 
\noindent {\bf Proof.} For details, please refer to Appendix B. \qquad $\square$\\

\nop{
\noindent {\bf Proof.} From the lemma assumptions, we have
$UB\_Pr\{W\_val(o_l,t_l)\leq\epsilon\}< 1-\alpha$. Then  
\begin{align*}
&UB\_Pr\{W\_val(o_l,t_l)\leq\epsilon\}+UB\_Pr\{W\_val(o_l,t_l)>\epsilon\}\geq 1\\  &\Longleftrightarrow 1\leq UB\_Pr\{W\_val(o_l,t_l)\leq\epsilon\}+UB\_Pr\{W\_val(o_l,t_l)\\ &>\epsilon\}<UB\_Pr\{W\_val(o_l,t_l)>\epsilon\}+1-\alpha\\
&\Longleftrightarrow UB\_Pr\{W\_val(o_l,t_l)>\epsilon\} > \alpha.
\end{align*}

From Definition \ref{def:weather_obstacle}, the edge $e_i$ contains $o_l$ is an obstacle at timestamp $t_l$. \qquad $\square$
}

\noindent{\bf Discussions on How to Compute Probability Upper Bound:}
In order to compute probability upper bound, $UB\_Pr\{W\_val(o_l,t_l)\leq \epsilon\}$, for a point $o_l\in e_i$ from weather predictions provided at $v_j$ and $v_k$, we may encounter three cases.

\noindent {\bf Case 1.} First, we consider the case where values of weather conditions at both $v_j$ and $v_k$ are less than or equal to weather threshold $\epsilon$ at timestamp $t_l$ (i.e., $W\_val(v_j, t_l)\leq \epsilon$ and $W\_val(v_k, t_l)\leq \epsilon$). In this case, the probability upper bound of weather values at any object $o_l\in e_i$ can be computed as:
\begin{eqnarray}
UB\_Pr\{W\_val(o_l,t_l)\leq \epsilon \}=1.
\label{eq:case1}
\end{eqnarray} 

The reason for the probability upper bound in Eq.~(\ref{eq:case1}) is as follows. In Cases \circled{A} $\sim$ \circled{C}, as mentioned in Section \ref{sec:weather_est}, the estimated weather values $W\_val(o_l, t_l)$ are definitely smaller than or equal to $\epsilon$ (due to the conditions of Case 1). For Case \circled{D}, the estimated weather value may be smaller than or equal to $\epsilon$. Therefore, the upper bound, $UB\_Pr\{.\}$, of the probability $Pr\{W\_val(o_l, t_l)\leq \epsilon\}$ is equal to 1 (i.e., Eq.~(\ref{eq:case1}) holds).

\begin{figure}[t!]
\centering
\subfigure[][{\small Case 1}]{
\scalebox{0.55}[0.55]{\includegraphics{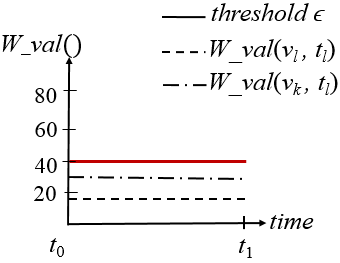}}\label{subfig:Wpruning1}
}
\subfigure[][{\small Case 2.1}]{
\scalebox{0.55}[0.55]{\includegraphics{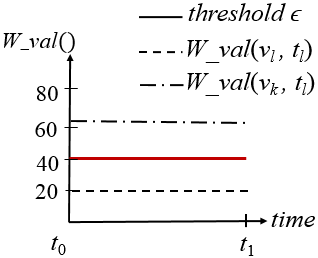}}\label{subfig:Wpruning2} 
}
\\
\subfigure[][{\small Case 3}]{                    
\scalebox{0.55}[0.55]{\includegraphics{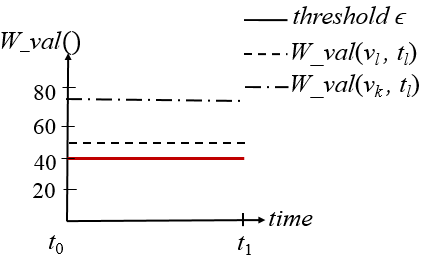}}\label{subfig:Wpruning3}
}
\caption{Illustration of weather-based pruning with wind speed values at $v_j$ and $v_k$ w.r.t. threshold $\epsilon$.}\label{fig:weatherpruning} \vspace{3ex}
\end{figure}

Figure \ref{subfig:Wpruning1} shows an example of Case 1, where the weather threshold $\epsilon$ is set to 40, and weather predictions at $v_j$ and $v_k$ are both less than $\epsilon$.

\noindent {\bf Case 2.} When $\epsilon$ is between weather values at two endpoints, $v_j$ and $v_k$, of edge $e_i$, we need to examine two sub-cases.

\underline{\it Case 2.1.} In the case where it holds that $W\_val(v_j,t_l)<\epsilon<W\_val(v_k,t_l)$, the probability upper pound is calculated as follows:
\begin{eqnarray}
&&UB\_Pr\{W\_val(o_l,t_l)\leq \epsilon\}
=p(v_k,t_l)\cdot p(v_j,t_l) \qquad\qquad\qquad \text{}\hspace{0.29cm}\text{// Case \circled{A}}\nonumber\\
&& \hspace{4.15cm}+ p(v_j,t_l)\cdot(1-p(v_k,t_l)) \quad \quad \hspace{0.39cm} \text{// Case \circled{C}}\nonumber\\ 
&& \hspace{4.15cm}+ (1-p(v_j, t_l))\cdot (1-p(v_k, t_l)) \text{  }\text{    //  Case \circled{D}}\nonumber\\
&& \hspace{4.15cm}= 1-p(v_k, t_l)\cdot (1-p(v_j, t_l)),\nonumber
\label{eq:case2.1}
\end{eqnarray}
where $p(v_j,t_l)$ (or $p(v_k, t_l)$) is the probability that the predicted weather is accurate for vertex $v_j$ (or $v_k$) at timestamp $t_l$.

In this scenario, as discussed in Section \ref{sec:weather_est}, the estimated weather values in Cases \circled{A}, \circled{C}, and \circled{D} may be smaller than or equal to $\epsilon$. Thus, we can overestimate their probabilities, and obtain the probability upper bound $UB\_Pr\{.\}$ by summing up appearance probabilities of these 3 cases. Please refer to an example of Case 2.1 in Figure \ref{subfig:Wpruning2}.

\underline{\it Case 2.2.} Similarly, in the case that $W\_val(v_k,t_l)<\epsilon<W\_val(v_j,t_l)$ holds, the probability upper bound for any point $o_l$ on edge $e_i$ can be given by:
\begin{eqnarray}
UB\_Pr\{W\_val(o_l,t_l)\leq \epsilon \}=1-p(v_j, t_l)\cdot (1-p(v_k, t_l)).
\label{eq:case2.2}
\end{eqnarray}

\noindent {\bf Case 3.} In the case that both predicted weather values at $v_j$ and $v_k$ are greater than the weather threshold $\epsilon$  (i.e., $W\_val(v_j, t_l) > \epsilon$ and $W\_val(v_k, t_l) > \epsilon$), the probability upper bound for any point $o_l$ on edge $e_i$ ($=v_jv_k$) at timestamp $t_l$ can be computed as follows:
\begin{eqnarray}
UB\_Pr\{W\_val(o_l,t_l)\leq \epsilon\}=(1-p(v_j,t_l))\cdot(1-p(v_k,t_l)).\hspace{-5ex}
\qquad \qquad \text{ // Case \circled{D}}
\label{eq:case3}
\end{eqnarray}

This is because the estimated weather values $W\_val(o_l,t_l)$ only has the chance to be greater than the weather threshold $\epsilon$ in Case \circled{D} (i.e., unknown value, as discussed in Section \ref{sec:weather_est}). Figure \ref{subfig:Wpruning3} shows an example of Case 3, where we set $\epsilon$ to 40, and weather values at both $v_j$ and $v_k$ are greater than threshold $\epsilon$.

\subsection{Traveling Time Pruning}
\label{subsec:travelingtimepuning}
In this subsection, we discuss the rationale of the \textit{traveling time pruning} method, which rules out paths, $Path$, with the traveling time greater than that of paths we have seen so far. For any path, $Path'$, we denoted $LB\_T(Path')$ and $UB\_T(Path')$ as lower and upper bounds of its traveling time, respectively, and {\it best-path-so-far} as a valid path we have seen so far from $src$ to $dst$ (satisfying all obstacle constraints) with the smallest traveling time. Then, our traveling time pruning method filters out paths, $Path$, whose lower bounds of the traveling times, $LB\_T(Path)$, are greater than the time upper bound $UB\_T$({\it best-path-so-far}) for path {\it best-path-so-far}. Intuitively, $Path$ cannot be the PRAO answer, if there exists a path, {\it best-path-so-far}, that has the traveling time smaller than that of $Path$.

\begin{lemma}  {\bf (Traveling Time Pruning)} Let {\it best-path-so-far} be the best path from $src$ to $dst$ that we have seen so far with the smallest traveling time. Then, any path, $Path$, can be safely pruned, if $UB\_T$({\it best-path-so-far}) $<$ $LB\_T(Path)$ holds. \label{lemma:lem3}
\end{lemma} 
\noindent {\bf Proof.} For details, please refer to Appendix C. \qquad $\square$\\

\nop{
\noindent {\bf Proof.} Let $UB\_T(\mbox{\it best-path-so-far})=(e_1.w+e_2.w+\ldots+e_{|\mbox{\it best-path-so-far}|}.w)$, where $e_1(=src\rightarrow v_j)$ and $e_{|\mbox{\it best-path-so-far}|}(=v_k\rightarrow dst)$. Suppose that, the lower bound traveling time of the path, $Path$, is $LB\_T(Path)=(e_1.w+e_2.w+\ldots+e_c.w)$, where $e_1(=src\rightarrow v_i)$ and $e_c(=v_f\rightarrow dst)$. 
 Since $UB\_T$({\it best-path-so-far})$< LB\_T(Path)$, by transitivity we have $T$({\it best-path-so-far})$< T(Path)$. The traveling time of the best-so-far path is less than the traveling time of $Path$. Thus, $Path$ can be safely pruned. \qquad $\square$
}

In order to enable the traveling time pruning, lower and upper bounds of the traveling time have to be efficiently obtained. In the sequel, we will discuss how to infer lower and upper bounds of the traveling time for a path, $Path'$ (based on the traveling times on edges $e_i\in Path'$).

\vspace{0.5ex}{\bf \noindent Discussion on How to Compute the Lower Bound of the Traveling Time $LB\_T(Path)$:} On road networks, the triangle inequality may not hold for the traveling time (due to variable vehicle speeds on roads), however, it holds for the shortest path distance.

As an example on road networks in Figure \ref{fig:EX}, it holds that $dist_N(v_2,v_4)\leq dist_N(v_2,v_5)+dist_N(v_5,v_4)$, but it may not be the case for the traveling time, that is, $T(v_2,v_4)> T(v_2,v_5)+T(v_5,v_4)$, which violates the triangle inequality. This may be due to the traffic jams on the road $v_2v_4$.

The traveling time on road networks can be obtained as the network distance divided by velocity. Thus, the lower bound, $LB\_T(Path)$, of the traveling time on any path, $Path$ ($=src \leadsto dst$), can be given by the lower bound of the shortest path distance, $LB\_dist_N(src, dst)$, divided by the maximum velocity on that path, that is,
\begin{eqnarray}
LB\_T(Path)= \frac{LB\_dist_N(src, dst)}{max\_vel(src \leadsto dst)},
\label{equ1}
\end{eqnarray} 
where and $max\_vel(src \leadsto dst)$ is the maximum velocity on the shortest path $src \leadsto dst$. Note that, here the velocity of each edge $e_i$ on $Path$ can be given by $\frac{e_i.len}{e_i.w}$.

\begin{figure}[t!]
  \centering
    \includegraphics[width=0.53\textwidth]{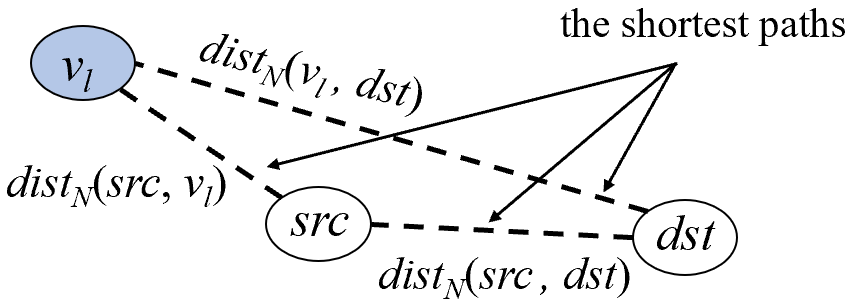}
     \caption{An example of computing the distance lower bound of the shortest path $src \leadsto dst$ on road networks.} 
     \label{fig:EX2}\vspace{3ex}
\end{figure}

\textit{\uline{The Computation of a Lower Bound of the Shortest Path Distance $LB\_dist_N(src, dst)$:}} We can obtain lower bounds of the shortest path distances on road networks, by utilizing the \textit{triangle inequality} property, as described in  \cite{goldberg2005computing}. In \cite{goldberg2005computing}, the distance bound is computed by choosing a small number of pivots, then computing and storing shortest path distance between all vertices and each of these pivots. By utilizing pivots, the lower bounds can be computed in constant time with the triangle inequality. In Figure \ref{fig:EX2}, assume that we have pre-computed $dist_N(src, v_l)$ and $dist_N(v_l, dst)$. Then, by the triangle inequality, we can obtain the lower bound of the shortest distance between $src$ and $dst$ as follows:
\begin{eqnarray}
\label{pivwithshortdistance}
dist_N(src, dst)&\geq& |dist_N(src, v_l)-dist_N(v_l, dst)|\label{equ3}\\ 
&=&LB\_dist_N(src, dst)\nonumber.
\end{eqnarray}

This way, we can utilize $d$ pivots (i.e., vertices on road networks, such as $v_l$ in the example of Figure \ref{fig:EX2}), and then offline pre-compute the shortest path distances from each vertex to these $d$ pivots. By using these $d$ pivots, we can apply Eq.~(\ref{pivwithshortdistance}) to compute a lower bound, $LB\_dist_N(src, dst)$, of the shortest path distance between any two vertices $src$ and $dst$. We will provide a cost model in Section \ref{subsec:piv_selection} to select good pivots on road networks.

{\bf \noindent Discussion on How to Calculate the Upper Bound of the Traveling Time $UB\_T$({\it best-path-so-far}):} In order to compute an upper bound of the traveling time, we only need to find a valid path, $Path$, from $src$ to $dst$ (satisfying all obstacle constraints), and treat the traveling time of this valid path as a time upper bound $UB\_T$({\it best-path-so-far}), which can be used for the traveling time pruning.

To achieve high pruning power, the selected path $Path$ ($= src \leadsto dst$) has to meet two requirements: (1) $Path$ has to be valid (i.e., $Path$ should not contain edges with ad-hoc keyword-based and weather-based obstacles), and (2) the traveling time on $Path$ should be close to the traveling time of the optimal path.

Due to the existence of ad-hoc obstacles, it is challenging to find a valid path from $src$ to $dst$. Prior works \cite{shahabi2002road} usually used pivots and the triangle inequality to compute time upper bounds, however, they did not consider ad-hocly given obstacles on roads. In contrast, some roads in our PRAO problem may not be available online due to ad-hoc obstacles, which may be passed by some offline pre-computed paths to pivots. Thus, we cannot utilize offline pre-computed distances via pivots and velocity bounds to calculate the traveling time upper bounds. Another challenge is that, due to many possible valid paths that connect $src$ and $dst$, it is not trivial how to efficiently choose a valid path with the traveling time close to that of the optimal path.



Inspired by the challenges above, in this paper, we will follow a greedy traversal algorithm to find a valid path, $Path$, that connects $src$ and $dst$. Then, the traveling time upper bound of the best-so-far path, {\it best-path-so-far}, from $src$ to $dst$ is equal to the minimum traveling time of $Path$ that we have encountered, that is,
\begin{eqnarray}
UB\_T({\it best\text{-}path\text{-}so\text{-}far})= \min_{\forall \text{ valid path } Path} T(Path),
\end{eqnarray}
\noindent where $T(Path)$ is the traveling time of a valid path $Path$.





\begin{figure}[t!]
  \centering
    \includegraphics[width=0.48\textwidth]{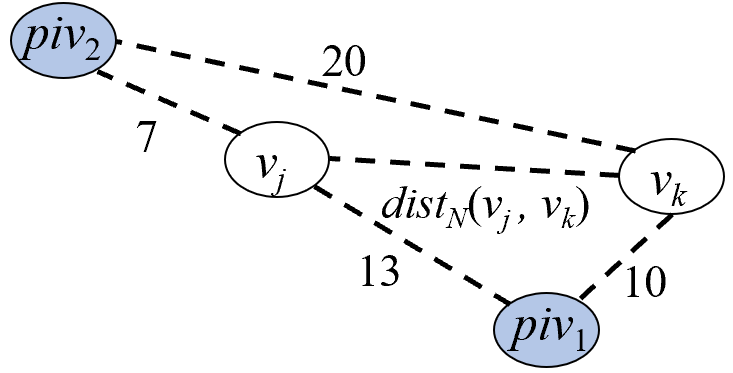}
     \caption{An example of computing the distance lower bound, with the help of 2 pivots, $piv_1$ and $piv_2$.} 
     \label{fig:costmodel}\vspace{3ex}
\end{figure}

\section{Offline Pre-Computations and Indexing}
\label{sec:query_processing}

In this section, we will illustrate offline pre-computation and indexing mechanisms over weather and keyword information on road networks, which can enable our proposed pruning methods and facilitate online PRAO query processing.

\subsection{Pivot Selection}
\label{subsec:piv_selection}
Next, we discuss our pivot selection algorithm in details including our proposed cost model and the difference with the landmark selection algorithm in \cite{goldberg2005computing}.




\noindent {\bf Cost Model for Pivot Selection.} As discussed in Section \ref{subsec:travelingtimepuning}, we can use the triangle inequality to estimate the lower bound of network distances between two vertices $v_j$ and $v_k$ via pivots. As an example, Figure \ref{fig:costmodel} shows two vertices, $v_j$ and $v_k$, and two pivots, $piv_1$ and $piv_2$, on road networks. We can offline pre-compute the shortest path distances from vertices $v_j$ and $v_k$ to these two pivots, that is, $dist_N(v_j, v_k)\geq |dist_N(v_j, piv_t)-dist_N(v_k, piv_t)|$, where $t=1, 2$. 

Intuitively, those pivots that result in large distance lower bound will give better pruning power. Thus, in Figure \ref{fig:costmodel}, $piv_2$ gives a tighter distance lower bound  (i.e., $13 = |7-20|$) than that of $piv_1$ (i.e., $3 = |13-10|$). 


From the discussion above, we will design a cost model below to evaluate the ``goodness'' of the selected pivots, based on the distance differences between pivots and pairs of vertices on road networks.
\begin{eqnarray}
\mathcal{C}= \sum_{\forall(v_i, v_j) \in V\times V} \max_{\forall piv_t \in \mathcal{S}_{piv}}|dist_N(v_i,piv_t)-dist_N(v_j,piv_t)|.
\label{costmodel}
\end{eqnarray}
\noindent where $\mathcal{S}_{piv}$ is a set of $d$ selected pivots.

Eq.~(\ref{costmodel}) computes the tightness, $\mathcal{C}$, of distance lower bounds (via $d$ pivots in a pivot set $\mathcal{S}_{piv}$), for all pairs of vertices $(v_i,v_j)$. Intuitively, larger $\mathcal{C}$ value will lead to higher pruning power. Therefore, our pivot selection algorithm aims to maximize the $\mathcal{C}$ value (i.e., based on our cost model) during the process of choosing pivots.


\noindent {\bf Pivot Selection Algorithm.} Algorithm \ref{piv_selection}, namely {\sf Obtain\_Pivots}, chooses a set, $\mathcal{S}_{piv}$, of $d$ pivots, in light of our proposed cost model (i.e., maximizing $\mathcal{C}$ in Eq.~(\ref{costmodel})). We first initialize two parameters, $global\_cost$ and $\mathcal{S}_{piv}$, which store the global $\mathcal{C}$ value and pivot set, respectively (line 1). Initially, our algorithm selects $d$ random pivots, forming a pivot set $\mathcal{P}$, and then evaluates the $\mathcal{C}$ value (recorded by $local\_cost$) with respect to $\mathcal{P}$ (lines 3-5).
Next, for $swap\_iter$ times, we will randomly pick a pivot $piv_i$ from $\mathcal{P}$ and a non-pivot $v_j\in V-\mathcal{P}$, and swap their roles so that a new pivot set $\mathcal{P}^{new}$ is obtained (lines 6-8). We evaluate the new cost, $\mathcal{C}^{new}$, over the updated pivot set $\mathcal{P}^{new}$ (line 9), and compare it with $local\_cost$ (lines 10). If it holds that $\mathcal{C}^{new} > local\_cost$, then we will accept new pivot set $\mathcal{P}^{new}$ and update $local\_cost$ (lines 11-12). The process of finding better pivots repeats for $swap\_iter$ times (lines 6-12). 

In order to avoid the locally optimal solution, we will execute the pseudo code above for $global\_iter$ times, by selecting different initial pivot sets (line 2). For each round, we will update the global pivot set $\mathcal{S}_{piv}$, if it holds that $local\_cost>global\_cost$ (lines 13-15). Finally, the algorithm returns the selected pivot set $\mathcal{S}_{piv}$ (line 16).



\noindent {\bf Discussions on the Number, $d$, of Pivots.} In order to decide an appropriate number of pivots, we will execute our pivot selection algorithm with different numbers of pivots, where $d = 1, 2, ...$. Intuitively, with more and more pivots, the increase of the pruning power (or the $\mathcal{C}$ value) will become smaller. We will set parameter $d$, such that the difference of $\mathcal{C}$ values with $d$ and $(d+1)$ pivots is below a user-specified threshold.

{
\noindent {\bf Discussions on the Differences Between Pivot Selection and Landmark Selection.} Note that, pivots used in our work are similar to the landmarks in \cite{goldberg2005computing}. Nevertheless, the criterion of selecting pivots in our PRAO algorithm is based on our proposed cost model (i.e., Eq. (\ref{costmodel})), which is different from the heuristic-based landmark selection algorithm. Specifically, landmarks in \cite{goldberg2005computing} are selected in order to provide tight distance bounds in graphs. The landmark selection algorithms are based on different heuristics such as greedy landmark selection, using landmarks geometrically lying behind the destination, and so on. With optimization techniques, \cite{goldberg2005computing} removes a landmark and replaces it with the best candidate landmark (with the highest score w.r.t. distance lower bounds).
In contrast, our work designs a specific cost model (as given in Eq. (\ref{costmodel})), which has a different goal (i.e., score) from the one for the landmark selection.
Furthermore, our pivot selection algorithm is in light of our proposed cost model, and aims to obtain the best pivot set with the highest $\mathcal{C}$ value (in Eq. (\ref{costmodel})) by avoiding local optimality. Thus, our cost-model-based pivot selection algorithm is different from heuristic-based landmark selection algorithm. 
}

\begin{algorithm}[t!]
\footnotesize
    \KwIn{road network $G=(V,E)$ and the number of pivots $d$}
    \KwOut{the set of pivots $\mathcal{S}_{piv}$}
    set $global\_{cost}=-\infty$, $\mathcal{S}_{piv}=\emptyset$
    
    \For{$a = 1$ to $global\_iter$}
    {
        randomly select $d$ pivots from $V$ and form a pivot set $\mathcal{P}$\\
        evaluate the cost function $\mathcal{C}$ of $\mathcal{P}$\\ set $local\_cost=\mathcal{C}$\\
        \For{$b = 1$ to $swap\_iter$}{
            select a random pivot $piv_i\in \mathcal{P}$\\
            randomly choose a non-pivot $v_j \in (V-\mathcal{P})$
            $\mathcal{P}^{new}= \mathcal{P}-\{piv_i\}+\{v_j\}$\\
            evaluate the cost function $\mathcal{C}^{new}$ w.r.t. $\mathcal{P}^{new}$\\
            \If {$\mathcal{C}^{new}> local\_cost$}{
                $local\_cost=\mathcal{C}^{new}$\\
                $\mathcal{P}=\mathcal{P}^{new}$
            }
        }
        \If{$local\_cost>global\_cost$}{
            $\mathcal{S}_{piv}=\mathcal{P}$\\
            $global\_cost=local\_cost$
        }
    }
 \Return $\mathcal{S}_{piv}$
    \caption{\sf Obtain\_Pivots}
    \label{piv_selection}\vspace{3ex}
\end{algorithm}

\subsection{Pre-Computations}
\label{Pre-Computations}
In this subsection, we discuss how to choose pivots and the data pre-computation.

\noindent{\bf{{Pre-Computation of the Shortest Path Distances with Pivots:}}}
Once we select pivots, we can pre-compute distances between all vertices $v_j\in V$ and pivots $piv_i$ (for $1\leq i\leq d$).
\begin{eqnarray}
dist_N(v_j, piv_i)= \sum_{\forall e_i \in Path(v_j\leadsto piv_i)}{e_i.len}.
\end{eqnarray}

The pre-computed distances with pivots are used to compute lower bounds of the traveling time, as discussed in Section \ref{subsec:travelingtimepuning}.

\noindent{\bf{Pre-Computation of the Probability Upper Bounds of Weather Conditions:}}
We pre-compute probability upper bounds of weather conditions, as well as weather values at endpoints of edges, as mentioned in Section \ref{sec:pruning}. The pre-computed data will be indexed in a tree index for facilitating online PRAO query processing. Specifically, given a user-specified weather threshold $\epsilon$ and a probabilistic threshold $\alpha$, we will utilize the pre-computed data to online compute the probability upper bound, based on Cases 1$sim$3 discussed in Section \ref{Weather-basedPruning}, which can be used to enable the weather-based pruning.

\subsection{Indexing}
\label{subsec:indexing}
In this subsection, we discuss the indexing mechanism over road networks, which can facilitate efficient PRAO query processing. In particular, we use an R$^*$-tree \cite{beckmann1990r}, denoted as $\mathcal{I}$, to index road network $G$ and its pre-computed data. To construct the R$^*$-tree, we first partition road networks (graphs) $G$ into subgraphs (each with edges spatially close to each other), and treat each subgraph as a leaf node, $N$, of the R$^*$-tree, which contains edges $e_i$ in the subgraph, represented by \textit{minimum bounding rectangles} (MBRs). Then, we recursively group MBRs of leaf or non-leaf nodes until we obtain a root of the R$^*$-tree. 


\vspace{0.5ex}\noindent{\bf Leaf Nodes:} Each edge $e_i$ ($=v_jv_k$) in the leaf node $N$ of index $\mathcal{I}$ is represented by an MBR, which minimally bounds all spatial locations in edge $e_i$. Moreover, edge $e_i$ is associated with two vectors of the pre-computed shortest path distances from the two endpoints of the edge $e_i$ to pivots, that is, $(dist_N(v_j,piv_1), dist_N(v_j,piv_2), \dots, dist_N(v_j,piv_d))$ and $(dist_N(v_k,piv_1), dist_N(v_k,piv_2), \dots,
dist_N(v_k,piv_d))$, the traveling time, $e_i.w$, on edge $e_i$, and a bitmap, $e_i.K$, that stores keywords associated with $e_i$. 

We also store the functions of weather predictions at endpoints $v_j$ and $v_k$ of edge $e_i$, that is, functions $W\_val(v_j, t)$ and $W\_val(v_k, t)$ at time $t$, respectively. In addition, for any edge $e_i$ $(=v_jv_k)$, we dynamically keep lower/upper bounds of weather predictions at edges $e_i$, denoted as $lb\_W\_val(e_i, t)$ and $ub\_W\_val(e_i, t)$, respectively, at timestamp $t$ in a future period of time. That is, we have:
\begin{eqnarray}
lb\_W\_val(e_i, t)=\min\{W\_val(v_j, t), W\_val(v_k, t)\},\\
ub\_W\_val(e_i, t)=\max\{W\_val(v_j, t), W\_val(v_k, t)\}.
\end{eqnarray}




\vspace{0.5ex}\noindent{\bf Non-Leaf Nodes:} Each non-leaf node, $N_p$, of index $\mathcal{I}$ contains multiple entries, denoted as $N_c$, each of which is represented by an MBR and minimally bounds edges in its subtree. Moreover, for each entry $N_c$, we also store minimum/maximum traveling times for any edge $e_i$ under $N_c$. 
\begin{eqnarray}
lb\_t(N_c) &=& \min_{\forall e_i\in N_c} e_i.w,\label{eq:node_lb_t}\\
ub\_t(N_c) &=& \max_{\forall e_i\in N_c} e_i.w.\label{eq:node_ub_t}
\end{eqnarray}

\nop{

Furthermore, entry $N_c$ stores lower/upper bounds of the shortest path distances from edges $e_i$ under entry $N_c$ to pivots $piv_j$ (for $1\leq j\leq d$):
\begin{eqnarray}
LB\_dist_N(N_c, piv_j)= \min_{\forall e_i\in N_c}\{dist_N(e_i, piv_j)\},\label{LB_nodes}\\
UB\_dist_N(N_c, piv_j)= \max_{\forall e_i\in N_c}\{dist_N(e_i, piv_j)\}.\label{UB_nodes}
\end{eqnarray}
We also store minimum/maximum velocities under entry $N_c$:
\begin{eqnarray}
min\_vel(N_c)= \min_{\forall e_i\in N_c}{\{e_i.vel\}},\\
max\_vel(N_c)= \max_{\forall e_i\in N_c}{\{e_i.vel\}}.
\end{eqnarray}

}

Furthermore, each entry $N_c\in \mathcal{I}$ is associated with a summary, $N_c.K$, of keywords shared by all edges under entry $N_c$, that is,
\begin{eqnarray}
N_c.K= \bigwedge_{\forall e_i\in N_c}{e_i.K}. 
\end{eqnarray}

We also dynamically maintain the upper/lower bounds of weather values over time for edges $e_i$ under node $N_c$, that is,
\begin{eqnarray}
lb\_W\_val(N_c, t)&=&\min_{\forall e_i\in N_c}\left\{lb\_W\_val(e_i, t)\right\},\nonumber\\
ub\_W\_val(N_c, t)&=&\max_{\forall e_i\in N_c}\left\{ub\_W\_val(e_i, t)\right\}; \nonumber
\end{eqnarray}
and probability upper bound of weather conditions for the 4 possible cases (as mentioned in Section \ref{sec:weather_est}):
\begin{eqnarray}
UB\_Pr\{N_c\}= \max_{\forall e_i\in N_c \wedge o_l\in e_i} \{UB\_Pr\{W\_val(o_l,t_l)\}\}.
\label{eq4}
\end{eqnarray}



\vspace{0.5ex}\noindent{\bf Dynamic Maintenance of Weather Data:}  In order to support efficient updates of weather data, for each entry of leaf/non-leaf nodes in the R$^*$-tree, we store a pointer pointing to a (space-efficient) circular array that contains a sliding window of weather prediction data in a future period of time (e.g., in the next 24 hours). Circular arrays are organized by a tree structure, the same as the structure of R$^*$-tree, which can be used for weather data maintenance.

\begin{figure}[t!]
  \centering\vspace{-2ex}
    \includegraphics[width=0.7\textwidth]{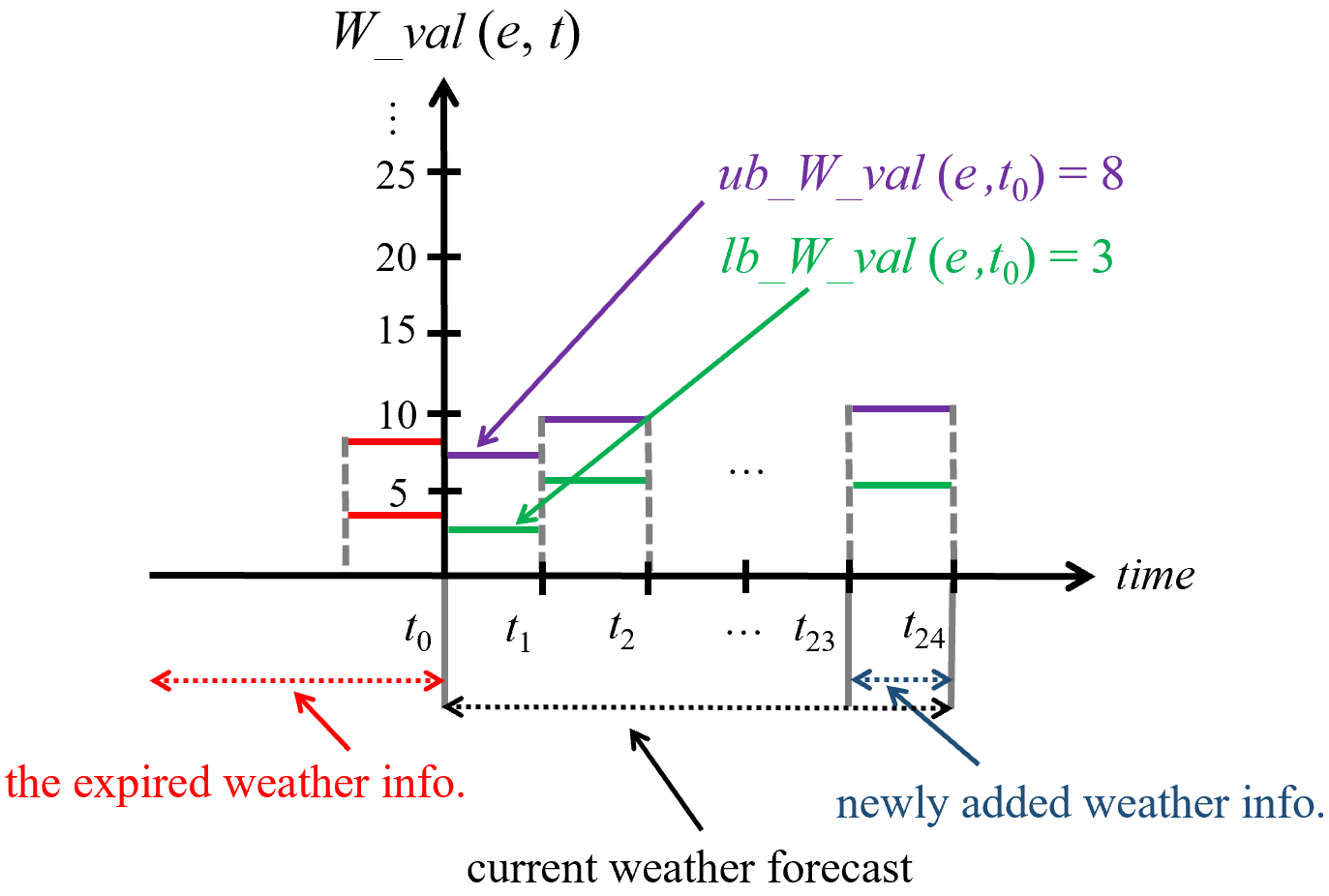}\vspace{-1ex}
     \caption{Dynamic update of weather forecast information for an edge on road networks (or an index entry). } 
     \label{fig:Wforecast}\vspace{1ex}
\end{figure}

As shown in Figure \ref{fig:Wforecast}, we maintain weather forecast information (e.g., weather lower/upper bounds and probability upper bounds) on edges or entries in the index in a streaming manner. When new weather prediction data on edges arrive, we will replace the expired weather data on edges with the new ones in circular arrays, and aggregate/update the weather information for circular arrays of index entries in a bottom-up manner. 

Note that, the R$^*$-tree index only needs to be offline constructed once, whereas the weather forecasting data are periodically maintained (e.g., every hour). We will show later in Section \ref{subsec:PRAO_exper_efficiency} that the time cost to dynamically update the weather data is low (e.g., about 0.01 $sec$, as will be confirmed in Figure \ref{fig:conVSupdate}).

\vspace{0.5ex}\noindent{\bf Auxiliary Synopses:} For the space efficiency, we use bitmap to store a summary of keywords in leaf or non-leaf nodes. On the leaf level, all possible keywords in $e_i.K$ are hashed into a bitmap. If some keyword on an edge $e_i$ is hashed to a bit in the bitmap, then this bit is set to 1; otherwise, the bit is set to 0. For leaf node $N$, we use the AND operator to aggregate all bitmaps of edges under $N$ and obtain a bitmap for leaf node $N$. Similarly, for non-leaf node $N_p$, we can also compute a bitmap by applying the AND operator over bitmaps of its children $N_c$.


To facilitate the path expansion during the index traversal, on each level of the R$^*$-tree, we will also maintain a \textit{connection graph}, where each vertex of this graph corresponds to a node entry, $N_i$, on this level, and each edge, $N_iN_j$, of the connection graph indicates that there are two road segments under two entries, $N_i$ and $N_j$, respectively, sharing the same endpoint (i.e., connected via an intersection point). 

In addition, in order to quickly look up index nodes containing a specific edge (e.g., the one containing source $src$) in road networks, we encode edges, as well as index nodes containing these edges, into a hash file, $Hash$.    

\subsection{Index Node Pruning}
\label{nodespruning}
In this subsection, we discuss the rational of index node pruning and illustrate the pruning of candidate paths that involve tree nodes $N_p$ in the index $\mathcal{I}$ using pruning methods discussed in Section \ref{section4}. {
In real-world scenarios, road-network obstacles (e.g., high temperature, high wind speed, etc.) may have an impact on an entire region. In such a case, all roads in this region that violate the weather constraints can be safely pruned. In other words, paths passing by these roads (with ad-hoc weather-based obstacles) can be safely pruned. Similarly, in the R$^*$-tree index, each tree node $N$ can be considered as a region containing multiple spatially close roads, associated with a summary of weather and keyword information in this region (as discussed in Section \ref{subsec:indexing}). Node $N$ can be directly pruned, if all edges under this node can be safely pruned (w.r.t. weather or keywords). This way, we can effectively prune an index node representing a spatial region that does not satisfy the constraints of weather conditions or keywords.

}


\vspace{0.5ex}{\noindent\bf Keyword-based Pruning:}  When we access a node $N_p\in \mathcal{I}$, we can check whether or not its child entries $N_c$ contain keyword-based obstacles. Specifically, each child entry $N_c$ is associated with a bitmap, $N_c.K$, which contains keyword information under node $N_c$. To check if node $N_c$ contains any obstacle keyword $s\in S$, we only need to verify if the bit corresponding to $s$ in bitmap $N_i.K$ is ``1''. If the answer is yes, it indicates that all edges under node $N_c$ contain the obstacle keyword $s$. Thus, none of edges under node $N_c$ can appear in the path of our PRAO query answer (due to the existence of the obstacle keyword), and entry $N_c$ can be safely pruned.


\vspace{0.5ex}{\noindent\bf Weather-based Pruning:} Each node entry $N_c \in \mathcal{I}$ stores an probability upper bound of the weather condition $UB\_Pr\{N_c\}$ (as given in Eq.~(\ref{eq4})). Intuitively, if this probability upper bound is less than $1-\alpha$ (i.e., $UB\_Pr\{N_c\}<1-\alpha$), then node $N_c$ can be safely pruned, where $\alpha$ is the probability threshold. This is because any object $o_l$ on  edges $e_i$ under node $N_c$ satisfies the following inequality: $$Pr\{W\_val(o_l,t_l)<\epsilon\}\leq UB\_Pr\{N_c\}.$$ Thus, any edge under $N_c$ cannot appear in our PRAO query answer, and node $N_c$ can be safely pruned. 



\vspace{0.5ex}{\noindent\bf Traveling Time Pruning:} We compute lower/upper bounds of traveling time for any edge under a node $N_c \in \mathcal{I}$. The bounds of traveling time of nodes can be used to compute lower/upper bounds of traveling time of paths, $Path = N_{src} \to N_1 \to \dots \to N_{dst}$ in the index $\mathcal{I}$.

Specifically, for any node $N_c$ in path $Path$, we can obtain the time lower/upper bounds, $LB\_T(N_c)$ and $UB\_T(N_c)$, of an edge under $N_c$, given by Eqs.~(\ref{eq:node_lb_t}) and (\ref{eq:node_ub_t}), respectively.

The only issue we need to address is the time lower/upper bounds for the first or last edge in a node $N_{src}$ of $Path$ (i.e., the one containing $src$ or $dst$). Without loss of generality, consider the first road segment $e_{src}$ in node $N_{src}$ containing $src$ on $Path$. Then, $LB\_T(N_{src}) = 0$ and $UB\_T(N_{src}) = ub\_t(N_{src})$. The case of node $N_{dst}$ is similar and thus omitted.

Based on Eqs.~(\ref{eq:node_lb_t}) and (\ref{eq:node_ub_t}), we can obtain the time lower bound, $LB\_T(Path)$, of a path $Path$ (even if this path does not reach the destination):
\begin{eqnarray}
LB\_T(Path) &=& \sum_{\forall N_c\in Path} lb\_t(N_c).\label{eq:path_lb_t}
\end{eqnarray}

Furthermore, when a path $Path$ is valid (i.e., satisfying keyword and weather constraints) and complete (i.e., reaching $dst$), we can obtain its time upper bound, $UB\_T(Path)$, as follows:
\begin{eqnarray}
UB\_T(Path) &=& \sum_{\forall N_c\in Path} ub\_t(N_c).\label{eq:path_ub_t}
\end{eqnarray}

\nop{

{\textit{\uline{Case 1: When $src$ and $dst$ Reside in the Same Node of the Index $\mathcal{I}$.}}} In this case, we utilize the triangle inequality to compute lower/upper bounds of distances as follows:
\begin{eqnarray}
&&|dist_N(src,piv_j)-dist_N(dst,piv_j)|\label{nodesenquality}\\
&\leq& dist_N(src,dst)\leq (dist_N(src,piv_j)+dist_N(dst,piv_j)).\nonumber
\end{eqnarray}

Based on Eq.~(\ref{nodesenquality}), we can 
estimate distance upper/lower bounds of node $N_p$ that contains $src$ and $dst$.
\begin{eqnarray}
&&lb\_dist_N(N_p)\label{lb_dist}\\
&=& \max_{j=1}^{d}{\{|dist_N(src, piv_j)-dist_N(dst, piv_j)|\}}\nonumber
\\
&&ub\_dist_N(N_p)\label{ub_dist}\\
&=& \min_{j=1}^{d}\{{dist_N(src, piv_j)+dist_N(dst, piv_j)\}}\nonumber
\end{eqnarray}

The lower/upper bounds of traveling time on the node $N_p$ can be computed as follows:
\begin{eqnarray}
LB\_T(N_p)=\frac{lb\_dist_N(N_p)}{max\_vel(N_p)},\label{nodelowerbound}\\
UB\_T(N_p)= \frac{ub\_dist_N(N_p)}{min\_vel(N_p)},\label{nodeupperbound}
\end{eqnarray}
where $ub\_dist_N(N_p)$ and $lb\_dist_N(N_p)$ are defined in Eqs.~(\ref{lb_dist}) and (\ref{ub_dist}), respectively, and $min\_vel(.)$ and $max\_vel(.)$ are given in Section \ref{Pre-Computations}.

{\textit{\uline{Case 2: When $src$ and $dst$ Reside in Two Different Node of the Index Tree $\mathcal{I}$.}}} In this scenario, we utilize the triangle inequality between nodes on the index tree $\mathcal{I}$ to compute lower/upper bounds of distances between nodes on the index tree. Assume that $src$ and $dst$ are reside in the node $N_i$ and $N_k$, respectively, then 
\begin{eqnarray}
&&lb\_dist_N(N_i,N_k)\label{lb_oftwonodes}\\
&=& \max_{j=1}^{d}{\{|LB\_dist_N(N_i,piv_j)-LB\_dist_N(N_k,piv_j)|\}},\nonumber\\
&&ub\_dist_N(N_i,N_k)\label{ub_oftwonodes}\\
&=& \min_{j=1}^{d}{\{UB\_dist_N(N_i,piv_j)+UB\_dist_N(N_k,piv_j)\}},\nonumber
\end{eqnarray}

in Eqs.~(\ref{lb_oftwonodes}) and (\ref{ub_oftwonodes}) values of $LB\_dist_N(.,.)$ and $UB\_dist_N(.,.)$ are pre-computed offline and stored in nodes as described in Eqs.~(\ref{LB_nodes}) and (\ref{UB_nodes}).

The bounds of travelling time between two nodes on index tree that contains the query points [$LB\_T(N_i,N_k), UB\_T(N_i,N_k)$] are computed as follows:
\begin{eqnarray}
LB\_T(N_i,N_k)=\frac{lb\_dist_N(N_i,N_k)}{\max\{max\_vel(N_i),max\_vel(N_k)\}},\label{LB_doublenodes}\\
UB\_T(N_i,N_k)=\frac{ub\_dist_N(N_i,N_k)}{\min\{min\_vel(N_i),min\_vel(N_k)\}},\label{UB_doublenodes}
\end{eqnarray}
where $lb\_dist_N(.,.)$ and $ub\_dist_N(.,.)$ are given in Eqs.~(\ref{lb_oftwonodes}) and (\ref{ub_oftwonodes}), respectively and $min\_vel(.)$ and $max\_vel(.)$ are pre-computed and stored at nodes as described in Section \ref{Pre-Computations}.

{\textit{\uline{Computation of Lower/Upper Bounds of Traveling Time of a Candidate Path on the Index Tree $\mathcal{I}$:}}}
The bounds of traveling time of a candidate path, $Path(N_i\leadsto N_k)$, on the index tree $\mathcal{I}$ can be computed as follows:
\begin{eqnarray}
&&LB\_T(Path)\\
&=& \left\{
\begin{array}{ll}
    LB\_T(N_p),& \mbox{if $Path$ contains only one node}\\ 
    LB\_T(N_i,N_k),& \mbox{if $Path$ contains two nodes}
\end{array}\right.\nonumber
\\
&&UB\_T(Path)\\
&=&\left\{
\begin{array}{ll}
    UB\_T(N_p),& \mbox{if $Path$ contains only one node}\\
    UB\_T(N_i,N_k),& \mbox{if $Path$ contains two nodes}
\end{array}\right. \nonumber
\end{eqnarray}
where $LB\_T(N_p)$ and $LB\_T(N_i,N_k)$ are defined in Eqs.~(\ref{nodelowerbound}) and (\ref{LB_doublenodes}), respectively, and $UB\_T(N_p)$ and $UB\_T(N_i,N_k)$ are given in Eqs.~(\ref{nodeupperbound}) and (\ref{UB_doublenodes}), respectively.

}

\nop{

{\color{ violet}
\subsection{Dynamic Update of Weather Information}
\label{subsec:MaintainingWeather}
To ensure a safe trip, the weather information has to be up to date. In this subsection, we propose an efficient method that dynamically updates aggregated weather information on the indexing tree nodes and the road network edges as well. 

As we noted earlier in subsection \ref{subsec:indexing}, indexing tree nodes hold a summary of weather data, probability of weather data, roads keyword, roads traveling time, etc.
Some of those parameters we refer as static in the sense that they are hard to change over time such as road keywords (a highway road). On the other hand, weather information parameters are changing overtime. The weather information expires and an up to date weather information may be collected.  

For simplicity let's assume that each edge of road network is represented by one low level node of the R$^*$-tree. For each edge of road network, we dynamically maintain weather forecast information for the next 24 hours. After an hour, one slot of weather information will be expired and a new weather slot will be added to maintain 24 hours of up to date weather information, as shown in Figure \ref{fig:Wforecast}. Furthermore, we maintain upper and lower bounds of up to date weather forecast information.

Assume that we can efficiently store the data from Figure \ref{fig:Wforecast} as well as the summary (upper and lower bounds) of weather data for each edge of road network. After a new weather slot has been added the weather update will be triggered in a bottom up manner.
The update process starts from the edge by updating its correspondent low level node of the indexing tree. After that, each low level node recursively updates summary information of its parent node until either no update required on an intermediate level node or the root is reached and updated (no more nodes to update).

Since it is highly inefficient to store data on the index, we suggest an auxiliary data structure that can store summary information for each node of index tree. Each node has a pointer that points to the appropriate location in the data structure where its summary information is located, as showing in Figure \ref{fig:treeUpdate}. The auxiliary data structure can be used directly to update information on the index.

Our proposed method of index dynamical update is very efficient for four reasons; (1) Weather data updates and the aggregation of data on each edge can be done in a constant time $O(1)$; (2) Index tree's node can update aggregated data of its parent node in a constant time ($O(1)$); 
(3) The height of index (R$^*$-tree) usually small constants (e.g., 3-4, or smaller);
And (4) many updates will not propagate high into the tree since the newly added information not always a lower or upper bound of weather data.
Based on the argument above, we say that our indexing update mechanism is very efficient.
Furthermore, we extensively evaluates the dynamic update mechanism in Section \ref{sec:exper}. 

It is worth mentioning that our proposed update mechanism can further efficiently update the static road network parameters such as traveling time or keyword parameters in case an update needed.

\begin{figure}[t!]
  \centering
    \includegraphics[width=0.48\textwidth]{treeUpdate}
     \caption{Dynamic update flow of the index tree and the auxiliary data structure.} 
     \label{fig:treeUpdate}\vspace{3ex}
\end{figure}

}

\nop{
To ensure a safe trip, the weather information has to be up to date. In this subsection, we propose an efficient offline method that dynamically updates aggregated weather information on the indexing tree. 

For simplicity, just for now we assume that the weather data is on hand and can be obtained in a constant time. Assume that we can obtain lower and upper values of weather condition and probability for each site. Note that each site is represented by a lower level node on the R$^*$-tree, as shown in Figure \ref{fig:update1}.
After aggregated weather information and probabilities on a indexing tree's node expires, the weather update on this node can be triggered. The lower and upper values of weather information will propagate up into the tree to update lower and upper values of weather information until either no more nodes to update or no update occurs on an intermediate node of the indexing tree (the new weather data do not update weather information on the parent node).

In order to achieve dynamic update, we place a timer, $timer$, on each node of the lower level of the index tree (R$^*$-tree index). The timer detects weather data expiration and triggers dynamic data update for the corresponding nodes of tree index. 
For each node on the index tree, only four values will be dynamically updated overtime, that are upper and lower values of weather condition ($UB\_W$, $LB\_W$, respectively) and upper and lower values of probability of weather condition ($UB\_Pr$, $LB\_Pr$, respectively) of sits represented by the tree nodes. The update process is shown in Algorithm \ref{alg.dynamic_update}. 

One bottleneck of dynamic update that we propose is the finding of aggregated upper and lower values of weather data and the probability of weather date. To solve this issue, we propose a dynamic data structure that dynamically keeps those aforementioned four values updated for each site.
whenever an update is triggered, the index's nodes can use the aforementioned data structure to update the aggregated values ($LB\_Pr, UB\_Pr, LB\_W, UB\_W$).

{\sf Dynamic\_update} (Algorithm \ref{alg.dynamic_update}) is very efficient for three reasons; (1) weather updates for sites occur very infrequently (every 1440 minutes); (2) the update in each node can be done in constant time; And (3) the height of index (R$^*$-tree) usually small constants (e.g., 3-4, or smaller). Based on the argument above, we say that our indexing update mechanism is very efficient and Figure \ref{fig:update1} shows the indexing tree with its content and the proposed data structure. 
It is worth mentioning that Algorithm \ref{alg.dynamic_update} can further efficiently update other road network parameters such as traveling time. In case of an update occurs on traveling time on one of the roads, the R$^*$-tree index does not need to be rebuilt again as for the four aforementioned parameters, it can update itself accordingly with minor changes to Algorithm \ref{alg.dynamic_update}.

\begin{algorithm}[t!]
    \KwIn{R$^*$-tree index}
    \KwOut{updated index}
    set $update =$ true \\
   \For{each node $N_l$ in the lower level of index tree}
    {\If{$timer=0$}
    {$timer=1440m$\\ 
    \While{$N_l \neq $ null And $update \neq $ false}
    { 
    {\If{$N_l$ is a lower level node Or any of parameters ($LB\_Pr, UB\_Pr, LB\_W, UB\_W$) needs update}
    {update parameters $LB\_Pr, UB\_Pr, LB\_W, UB\_W$\\
    $update =$ true}
    \Else{$update=$ false}
    $N_l=parent(N_l)$}}}}

    \caption{\sf Dynamic\_update}
    \label{alg.dynamic_update}\vspace{3ex}
\end{algorithm}

\begin{figure}[t!]
  \centering
    \includegraphics[width=0.4\textwidth]{dynamic}
     \caption{R$^*$-tree index with nodes, aggregated vales, and auxiliary data structure.} 
     \label{fig:update1}\vspace{3ex}
\end{figure}

}

\nop{
To ensure a safe trip, the weather information has to be up to date. In this subsection, we propose an efficient offline method that dynamically updates aggregated weather information on the indexing tree and of the network objects.

The process of checking the expiration of weather information and update it (in case the information was out of date) at query time is extremely time consuming; that is because the weather data is requested from an online server (e.g., Dark Sky).

Based on the argument above, instead of collecting weather data at query time we suggest an auxiliary data structure that collects the updated weather information separately from the query processing algorithm. The auxiliary data structure can be utilized to update the expired weather data on the index tree and road network objects.

We suggest a timing approach such that after a known period of time the weather information will be expired and the weather information update will be triggered. On the lower level of the index tree, we set a timer, t, (shown in Figure \ref{fig:update1}) that can trigger an update for weather information after a specified period of time. In such a case no weather updates are needed at query time.

The auxiliary data structure is very critical for boosting the weather information update's speed. Independent from the query processing algorithm, the auxiliary data structure constantly updates weather information by connecting to the server. The network objects are connected directly to the auxiliary data structure to get the updated weather information in a constant time. The auxiliary data structure also stores four values for each site, that are maximum and minimum value for weather condition, and maximum and minimum probability for weather condition. At update time, if any of the four values has changes, then an update for the four values has to be triggered on the tree node that initiate the update.

In our work, we set the timer to 48 hours, meaning we collect a correct data that are valid for two days. After that, the data will expire and the update will be triggered. For the auxiliary data structure, we use a two dimensional array to store the weather information for a constant time access as shown in Figure \ref{fig:update1}.

\begin{figure}[t!]
  \centering
    \includegraphics[width=0.5\textwidth]{UP}
     \caption{Weather information update process.} 
     \label{fig:update1}\vspace{3ex}
\end{figure}
}
}

\begin{algorithm}[p!]\scriptsize
    \KwIn{a road network $G = (V,E)$, an index $\mathcal{I}$ over $G$, a source $src$, a destination $dst$, a set, $S$, of obstacle keywords, a weather threshold $\epsilon$, and a probabilistic threshold $\alpha$}
   \KwOut{a path, $Path$, that satisfies Definition \ref{def6}}
   initialize an empty queue, $\mathcal{Q}$, with entries in the form ($Path, level$)\\
   initialize an empty min-heap, $\mathcal{H}$, accepting ($key,Path$) pairs\\
   $\lambda\leftarrow +\infty$; $\mathcal{S}_{cand}\leftarrow \phi$$; level \leftarrow$ height( $\mathcal{I}$)\\
   
   obtain the entry $N_{src} \in root(\mathcal{I})$ that contains $src$\\
   
   
   \For{each path, Path = $N_{src}\rightarrow N_i$ $(\in root(\mathcal{I}))$}
   {
   apply keyword-based and weather-based pruning\label{line: pruning1}\\
   \If{$Path$ cannot be pruned}
   {
   $\mathcal{Q}\leftarrow (Path,level)$
   }
   }
   
   
   
   
    \For{each path, $(Path, level)$, in $\mathcal{Q}$}
    {
        \If{$level=-1$} {terminate the loop}
        $(Path, level) = pop\text{-}out(\mathcal{Q}$);\\
        substitute node entries in $Path$ with their children, and obtain paths $Path_c$\\
        
        \For{each $Path_c$}
        {
        expand $Path_c$ by increasing the length by 1, and obtain $Path_c'$\\
        apply keyword-based and weather-based pruning\\
        \If{$Path_c'$ cannot be pruned}
        {
        $\mathcal{Q}\leftarrow (Path_c',level-1)$
        }
        }


    }
     
    \If {$\mathcal{Q}$ is empty}
   {
    \Return $\emptyset$\\
   }
       \For{each $Path \in \mathcal{Q}$}
       {
       \If{$Path$ reaches $dst$}{
        \If{all edges, $e_i\in Path$, satisfy the weather condition, and $Path$ cannot be pruned by traveling time pruning}{
        $\mathcal{S}_{cand} \leftarrow \mathcal{S}_{cand} \cup \{Path\}$\\
        \If{$\lambda > UB\_T(Path)$}
       {
        $\lambda \leftarrow UB\_T(Path)$
       }
       }
       
       }\Else{insert the entry ($LB\_T(Path),Path$), into $\mathcal{H}$\\}
       }


\While{$\mathcal{H}$ is not empty \label{whileline}}
    {
        $(key,Path)=pop\text{-}out(\mathcal{H})$\label{extract}\\
       \If{$key > \lambda$ \label{ifkey}} 
       {
       terminate the loop\\
       }
       
        expand the path, $Path$, by increasing the length by 1, and obtain path, $Path'$\label{expand}\\

        \If {$Path'$ cannot be pruned by keyword-based, weather-based, and traveling time pruning}
        {
        \If{$Path'$ reaches $dst$ \label{ifreaches}}
       {
        $\mathcal{S}_{cand} \leftarrow \mathcal{S}_{cand} \cup \{Path'\}$\\
        \If{$\lambda > T(Path')$}
       {
        $\lambda \leftarrow T(Path')$\label{setlambda}
       }
       }\Else{ insert the entry, ($T(Path'),Path'$), into $\mathcal{H}$ \label{insertinheap}\\}
        }
    }
    refine candidate paths in $\mathcal{S}_{cand}$ and return the actual PRAO answer\label{refine}\\
    \caption{\sf PRAO-QP}
    \label{prao-qp}\vspace{2ex}
\end{algorithm}
\section{Processing of PRAO Queries}
\label{sec:QPalgorithm}
\subsection{The PRAO Query Procedure}

In this subsection, we will propose an efficient query processing approach to answer PRAO queries. Specifically, Algorithm \ref{prao-qp} illustrates the pseudo code of our PRAO query answering algorithm, namely {\sf PRAO-QP}, which follows the filter-and-refine framework. That is, {\sf PRAO-QP} first obtains candidate paths of PRAO answers by traversing the tree index $\mathcal{I}$ and applying our three pruning methods (keyword-based, weather-based, and traveling time pruning), stated in Section \ref{sec:pruning}. Then, we refine the retrieved candidate paths (via the index traversal), and return the actual PRAO answers.

\vspace{0.5ex}\noindent{\bf Initialization:} Specifically, in Algorithm \ref{prao-qp}, we first initialize an empty queue $\mathcal{Q}$ for traversing index $\mathcal{I}$, which accepts entries in the form of $(Path, level)$, where $Path$ is a path containing index nodes, and $level$ is the level of nodes in path $Path$ (line 1). We also keep a minimum heap, $\mathcal{H}$, with entries in the form $(key, Path, t)$ for path expansion after the index traversal, where $key$ stores the lower bound of the traveling time on a path $Path$ and $t$ is the timestamp that is the time spend on $Path$ added to the query time (line 2). In addition, we also set a traveling time threshold $\lambda$ to $+\infty$, maintain an empty candidate set $S_{cand}$, and let initial traversing level, $level$, be the height, $height(\mathcal{I})$, of the R$^*$-tree (line 3).

\noindent{\bf Index Traversal:} Next, we will obtain an entry, $N_{src}$ in root, $root(\mathcal{I})$, that contains the source $src$, and obtain all paths, $Path = N_{src}\rightarrow N_i$, of length 2 (note: each node in $Path$ represents one edge), which start from an edge in entry $N_{src}$ and pass the keyword-based and weather-based pruning, where $N_i\in root(\mathcal{I})$. We will then add these paths in the form $(Path, level)$ to queue $\mathcal{Q}$ (lines 4-8). 

In the sequel, we will use queue $\mathcal{Q}$ to traverse the R$^*$-tree in a breadth-first manner and retrieve candidate paths (lines 9-28). In particular, if the candidate path, $Path$, from queue $\mathcal{Q}$ contains entries on the edge level (i.e., $level=-1$), then we will terminate the loop (lines 9-11); otherwise, we will pop out a candidate path $(Path, level)$ from queue $\mathcal{Q}$, and substitute node entries in $Path$ with their children, which leads to paths, $Path_c$, on lower level (lines 12-13). Furthermore, for these paths $Path_c$, we will try to expand the path $Path_c$ (via \textit{connection graphs} mentioned in Section \ref{subsec:indexing}) by including one more (connected) node entry on level $(level-1)$, and obtain paths $Path_c'$ (lines 14-16). If the expanded path $Path_c'$ cannot be ruled out by keyword-based and weather-based pruning, then we will insert it into queue $\mathcal{Q}$ for further checking (lines 17-18). The loop terminates when queue $\mathcal{Q}$ is empty (line 9) or the index traversal reaches the edge level (lines 10-11).

When queue $\mathcal{Q}$ is empty, it indicates that we cannot find candidate paths, and thus return an empty query answer set $\emptyset$ (lines 19-20). Otherwise, we will prepare for expanding the length of candidate paths in $\mathcal{Q}$ (lines 21-28). If paths $Path$ in $\mathcal{Q}$ have reached destination $dst$, then we will check whether or not they satisfy the weather and traveling time pruning (lines 22-23). Those objects that cannot be pruned will be added to the candidate set $S_{cand}$, and threshold $\lambda$ will be updated with the smallest upper bound of the traveling time for candidate paths we have seen so far (discussed in Section \ref{subsec:travelingtimepuning}; lines 24-26). If the paths $Path \in \mathcal{Q}$ have not reached $dst$, then we will insert them into heap $\mathcal{H}$ in the form $(LB\_T(Path), Path)$, where $LB\_T(Path)$ is the time lower bound  for (partial) path $Path$ (as the path expansion is not completed; lines 27-28).

\vspace{0.5ex}\noindent{\bf Path Expansion:} Finally, we will use the heap $\mathcal{H}$ to further expand (partial) paths in it, and obtain complete paths from $src$ to $dst$ (lines 29-40). Specifically, each time we obtain an entry $(key, Path)$ from heap $\mathcal{H}$ with the minimum key (i.e., time lower bound; lines 29-30). Then, we try to expand the path $Path$ by increasing its length by 1, and obtain a path $Path'$ (lines 33). If path $Path'$ can pass the weather-based, keyword-based, and traveling time pruning and reaches destination $dst$, we will add it to the candidate set $S_{cand}$ and update threshold $\lambda$ (lines 34-38); otherwise (i.e., $Path'$ cannot be pruned and does not reach $dst$), we insert $(T(Path'), Path')$ into heap $\mathcal{H}$ for further expansion (lines 39-40). The path expansion loop repeats until heap $\mathcal{H}$ is empty (line 29) or the popped entry has $key$ value above threshold $\lambda$ (i.e., all the remaining paths in $\mathcal{H}$ cannot have smaller traveling time; lines 31-32).

\vspace{0.5ex}\noindent{\bf Candidate Path Refinement:} After the index traversal and path expansion, we can obtain a number of candidate paths in set $\mathcal{S}_{cand}$. Then, we can refine these paths by checking constraints of keywords, weather, and traveling times, and return actual PRAO answers (line 41).

{Next, we give a running example for the {\sf PRAO-QP} algorithm in Algorithm \ref{prao-qp}. Figure \ref{fig:PRAO_ex1} shows an example of road networks and its corresponding graph representation, where the areas within red ovals correspond to those regions with bad weather conditions and the dashed (red) edges indicate roads with ad-hoc keyword obstacles. Moreover, Figure \ref{fig:PRAO_ex2} shows the corresponding R$^*$-tree index of road networks in Figure \ref{fig:PRAO_ex1}. Below, we give the detailed steps for this running example.

\begin{minipage}[t]{0.9\linewidth}
\begin{itemize}
\linespread{0.5}\fontsize{10pt}{10pt}\selectfont
    {
    \item[-] The {\sf PRAO-QP} procedure starts by obtaining an entry $N_{src}$ that contains the $src$ edge, $N_1$,
    \item[-] obtain all paths, $Paths$,of length 2 that starts from $N_1$, that is, \{\{$N_1N_1$\},\{$N_1 N_2$\}\}
    \item[-] apply weather-based and keyword-based pruning on $Paths$
    \item[-] insert remaining paths into the queue $\mathcal{Q}$ in the form of ($Path, level$), $\mathcal{Q} <(\{ N_1 N_1\}, 1)(\{N_1 N_2\}, 1)>$
    \item[-] for all paths, $Path \in \mathcal{Q}$
    
    \subitem --extract $Path$ for $\mathcal{Q}$
    \subitem --substitute node entries in $Path$ with their children and expand the length of the path by 1,
    \subsubitem ---resulting path is: $\{ N_3N_4N_4\}$
    \subitem --apply weather-based and keyword-based pruning on the node level
    \subsubitem ---$Path$ $\{N_3N_4N_4\}$ will be directly pruned due to the existence of an unsafe node $N_4$\\
    \textit{**note that no path exists, to continue, we extract another path from the queue $\mathcal{Q}$}
    \subitem --extract $Path$ from $\mathcal{Q}$, $Path= \{N_1 N_2\}$
    \subitem --substitute node entries in $Path$ with their children and expand the length of the path by 1, 
    \subsubitem ---resulting paths are: $\{\{N_3N_4N_5\}, \{N_3N_4N_6\}, \{N_3N_4N_7\}, \{N_3N_5N_6\}, \{N_3N_5N_7\}, \{N_3N_6N_7\}\}$
    \subitem --apply weather-based and keyword-based pruning on the node level
    \subsubitem ---$Paths$ $\{N_3N_4N_5\}, \{N_3N_4N_6\}, \{N_3N_4N_7\}, \{N_3N_5N_6\}, \{N_3N_5N_7\}$ will be directly pruned due to the existence of unsafe nodes $N_4$ and $N_5$
    \subitem --insert the remaining path into the queue $\mathcal{Q}<(\{N_3N_6N_7\}, 0)>$
    \subitem --extract paths from the queue, $Path= \{N_3N_6N_7\}$
    \subitem --substitute node entries in $Path$ with their children and expand the length of the path by 1
    \subitem --compute upper bound of traveling time $UB\_T(Path)$
    \subsubitem ---resulting paths are:	$\{\{e_2, e_3, e_{13} , e_{26}\}, UB\_T(Path)\}$
    \subitem --insert the resulting paths into the heap $\mathcal{H}$
    \subitem --for all $Paths$ in the heap $\mathcal{H}$
        \subsubitem ---further expand partial paths by increase length by 1 and obtain $Path’$, apply weather-based, keyword-based, and traveling time pruning 
        \subsubitem ---insert pair ($Path’, T(Path’)$) in $\mathcal{H}$
        \subsubitem ---continue until either the heap is empty, or a valid path from $src$ to $dst$ is formed with traveling time greater than the upper bound of traveling time of all paths in the heap $\mathcal {H}$
}
\end{itemize}
\end{minipage}

\begin{figure}[t!]
\centering
\subfigure[][{\small Illustration of road network with its corresponding graph representation.}]{
\scalebox{0.3}[0.3]{\includegraphics{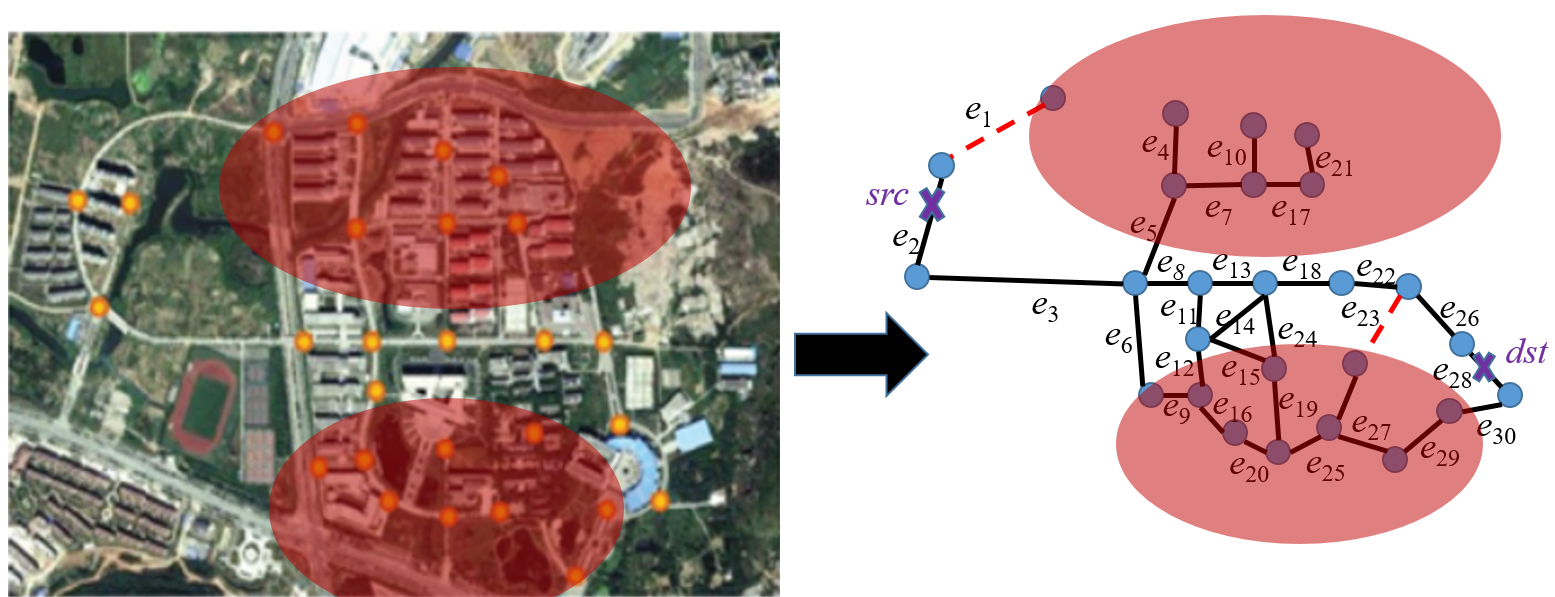}}\label{fig:PRAO_ex1}
}
\\
\subfigure[][{\small The R$^*$-tree representation of road network in (a).}]{
\scalebox{0.3}[0.3]{\includegraphics{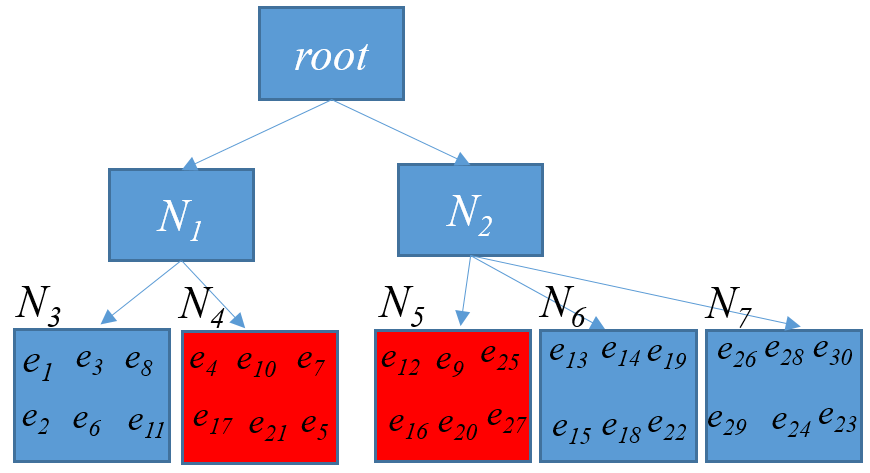}}\label{fig:PRAO_ex2} 
}
\caption{Illustration of road network and its corresponding graph representation and R$^*$-tree index.}\label{fig:PRAO_ex} \vspace{3ex}
\end{figure}

}

\subsection{Complexity Analysis}


In this subsection, we discuss the time complexity of the {\sf PRAO-QP} algorithm in Algorithm \ref{prao-qp}. The major time cost of Algorithm \ref{prao-qp} consists of two portions: index traversal (lines 1-28) and candidate path expansion (lines 29-41). 

Without loss of generality, assume that the \textit{connection graph} (mentioned in Section \ref{subsec:indexing}) for the $i$-th level of the R$^*$-tree index has the average degree $deg_i$. Then, during the index traversal, on the $i$-th level, the number of candidate paths $Path$ (starting from $N_{src}$ that contains $src$) can be given by $\sum_{j=1}^{h-i+1} (deg_i+1)^j \big(=\frac{(deg_i+1)^{h-i+1} -1}{deg_i} \big)$,
where $h$ is the height $height(\mathcal{I})$ of the R$^*$-tree. Moreover, in the worst case, we need to expand these candidate paths by including one more node. Thus, the traversal cost on the $i$-th level is given by $\frac{(deg_i+1)^{h-i+1} -1}{deg_i}\cdot (deg_i+1)$, and the total index traversal cost, $C_{index}$, is given by $\sum_{i=0}^h \big(\frac{(deg_i+1)^{h-i+1} -1}{deg_i}\cdot (deg_i+1)\big)$.

Furthermore, for candidate path expansion, there are totally about $(deg_0+1)^{h+1}$ candidate paths on the leaf level after the index traversal. Assume that the average degree of road network $G$ is denoted by $deg_G$, and the length of valid path from $src$ to $dst$ is represented by $\sigma$. Then, the total cost of path expansions, $C_{expansion}$, can be given by $(deg_0+1)^{h+1}\cdot deg_G^{\sigma-h-1}$ in the worst case.

To summarize, in the worst case (i.e., candidate paths cannot be pruned by any pruning strategies), the worst-case time complexity of Algorithm \ref{prao-qp} is given by $C_{index} + C_{expansion} = \sum_{i=0}^h \big(\frac{(deg_i+1)^{h-i+1} -1}{deg_i}\cdot (deg_i+1)\big) + (deg_0+1)^{h+1}\cdot deg_G^{\sigma-h-1}$, or $O\left(\sum_{i=0}^h (deg_i+1)^{h-i+1} + deg_0^{h+1}\cdot deg_G^{\sigma-h-1}\right)$. In practice, degrees $deg_i$ (or $deg_G$) of connection graphs (or road networks), and height, $h$, of the R$^*$-tree are usually small constants (e.g., 3-4, or smaller), and our pruning methods (via keyword, weather, and time constraints) are effective to prune candidate paths. Therefore, as confirmed by empirical studies in Section \ref{sec:exper}, the {\sf PRAO-QP} algorithm can achieve good performance in the amortized case, in terms of the CPU time and I/O cost.  



\nop{
The {\sf PRAO-QP} (Algorithm \ref{prao-qp}) starts by obtaining entry ($N_{src}\in root(\mathcal{I})$), (line 4), this step has a constant cost $O(1)$,  In line 5 of Algorithm \ref{prao-qp}, the cost is proportional to the number of paths of length 1, from $N_{src}$, that is the degree of $N_{src}$. The keyword-based, and weather-based pruning (line 6), has a constant time, discussed in Section 
\ref{sec:pruning}.
}

\nop{
Most of the time consuming work for our approach is done offline as we build R$^*$-tree that is $O(ElogE)$, where $E$ is the number of edges, select pivots (stated in subsection \ref{sub:piv_selection}), and summarize edge information in their corresponding tree nodes. The cost of summarizing edge information is corresponding to traversal of the R$^*$-tree, since we traverse the tree from down up to summarize edge information. The cost of summarizing step is $OElogE$.
The {\sf PRAO-QP} complexity comes in two part; (1) traversing the R$^*$-tree
}

\nop{
for all paths in $\mathcal{Q}$, {\sf PRAO-QP} extracts one path, $Path$, with its associated level, $level$, at a time. For each node $N\in Path$, substitute $N$ with its children nodes in the lower level ($level-1$) and obtain paths $Path_c$. Extend the new path, $Path_c$ by one to obtain $Path_c'$. At the same time of extending, we apply keyword-based and weather-based pruning to prune false alarms, paths that do not satisfy keyword or weather pruning. If $Path_c'$ survives pruning, it will be en-queued in the queue, $\mathcal{Q}$, with its associated level on the $\mathcal{I}$ tree, ($Path_c', level$). This process will continue until we reach to the leaf level where the objects (edges) are stored in the leaf level.

After reaching to the leaf level, if the queue, $\mathcal{Q}$, is empty, that means no path satisfies obstacles can be exist and we return $\emptyset$; Otherwise, we continue to the third step, that is we do more refinement to the paths by applying our third pruning method  that is traveling time pruning. In the third step, we extract a path, $Path$ from $\mathcal{Q}$ one a time. If the $Path$ reaches destination, then we apply weather-based and traveling time pruning. If $Path$ survived pruning, we add path to the candidate set $\mathcal{S}_{cand}$, and if $\lambda$ is greater than the upper bound of traveling time of $Path$, $UB\_T(Path)$, $\lambda$ is set to the upper bound of traveling time of $Path$. $\lambda$ will be utilized to pruning false alarms that is paths with high traveling time. In the case of we extract path, $Path$, from the queue, $\mathcal{Q}$, and $Path$ does not reach destination, then we compute lower bound traveling time of $Path$, that is $LB\_T(Path)$, and we insert $Path$ into min-heap, $\mathcal{H}$, with key equal to its lower bound of traveling time, ($LB\_T(Path), Path$).

Next, we extract paths from the min-heap, $\mathcal{H}$, (line \ref{extract}) with its associated key value. in line \ref{ifkey}, if $key$ is greater than $\lambda$, that means all paths in the min-heap have a traveling time greater than paths in the candidate set, $\mathcal{S}_{cand}$; We stop extracting paths and go to the refinement step (line \ref{refine}). In the case of the extracted path, $Path$, does not lead to a loop termination, we expand $Path$ by increasing its length by one edge and obtain $Path'$ (line \ref{expand}).
We apply keyword-based, weather-based, and traveling time pruning on $Path'$. If it can be pruned by any of pruning methods, we discard $Path'$.
Otherwise, if the path $Path'$ reaches destination, then we add $Path'$ to the candidate set, $\mathcal{S}_{cand}$, and if $\lambda$ is greater than the exact traveling time of $Path'$, then $\lambda$ is set to the exact traveling time of $Path'$, (lines \ref{ifreaches}-\ref{setlambda}). In the case of $Path'$ cannot be pruned and cannot reach destination, we insert $Path'$ in the min-heap, $\mathcal{H}$, with $key$
equal to the actual traveling time of $Path'$, line (line \ref{insertinheap}). We continue the process until one of the two condition applied: (1) the min-heap is empty, no more paths to extract from the heap; (2) condition in line (\ref{ifkey}) is satisfied, that means all paths remaining in the min-heap has a higher traveling time than paths in the candidate set $\mathcal{S}_{cand}$. 
}

\section{Experimental Evaluation}
\label{sec:exper}

\subsection{Experimental Settings}

\vspace{0.5ex}\noindent {\bf Real/synthetic Data Sets:} We test our PRAO query answering approaches on both real and synthetic data sets. Specifically, for real graph data, we use the California Road Network \cite{CA_network}, denoted as $CA$, which contains 21,048 road intersection points and 21,693 road segments. $CA$ is originally obtained from Digital Chart of the World Server and U.S. Geological Survey. Each vertex in $CA$ data set is represented by (longitude, latitude), and we obtain real weather forecast of each vertex from Dark Sky \cite{dark}. Moreover, we produce random integers within interval [0, 15] to simulate keywords associated with roads (edges), which may correspond to highway, city, uneven road, and so on. We also obtain the average traveling times on roads that are proportional to lengths of roads.

For synthetic data, we generate road networks (i.e., graphs) as follows. Specifically, we first generate random vertices in the 2D data space following either Uniform or Gaussian distribution. Then, we randomly connect vertices nearby through edges, such that all vertices are reachable in one single connected graph and the average degree of vertices is within $[3, 4]$. This way, with Uniform and Gaussian distributions of vertices, we can obtain two types of graphs, denoted as $Uni$ and $Gau$, respectively. For each vertex, we also generate random weather data, for example, wind speed within interval $[0, 100]$ and its forecasting accuracy within $(0, 1]$. Then, we also associate each edge $e_i$ with keywords in $e_i.K$, and the traveling time $e_i.w$.

We build indexes (including a variant of R$^*$-tree \cite{beckmann1990r} and auxiliary data structures) over each of real/synthetic road networks above (as mentioned in Section \ref{subsec:indexing}), where the page size is set to $4K$. For the pivot selection, we set the number of pivots to 5, based on the cost model as discussed in Section \ref{subsec:piv_selection}.




To evaluate the PRAO query performance, we generate 20 queries with source $src$ and destination $dst$, where the average length, $\sigma$, of valid paths between $src$ and $dst$ is set to 5, 8, 10 (default value), 15, 20, 25, or 30. 

\vspace{0.5ex}\noindent {\bf Measures:} We will report the CPU time and the I/O cost of our PRAO query processing approach, where the CPU time is the average time cost of conducting PRAO queries, and the I/O cost is defined by the number of node accesses during the index traversal.


{
\vspace{0.5ex}\noindent {\bf Competitor:} To the best of our knowledge, there are no prior works that answer path routing queries by considering both ad-hoc keyword-based and weather-based obstacles.
In this paper, we compare our PRAO algorithm with two baseline algorithms, denoted as $A^*$ and $fliterfirst$. Specifically,
the first baseline algorithm, $A^*$, is a variant of the $A^*$ algorithm \cite{Dijkstra59}. Note that, we cannot directly apply the $A^*$ algorithm, due to the existence of the ad-hoc keyword-based and weather-based obstacles. Thus, we adapt the $A^*$ algorithm to handle weather-based and keyword-based obstacles as follows. The $A^*$ algorithm traverses the graph in a breadth-first manner. Whenever encountering an edge, we check the edge to see if it contains weather-based and/or keyword-based obstacles. In case the edge violates the weather/keyword constraints, it will be pruned; otherwise, it will be inserted into the queue for further checking.

The second baseline algorithm, $filterfirst$, starts by filtering out all edges in road networks that contain ad-hoc over the remaining road networks (excluding those with obstacles)  weather-based and/or keyword-based obstacles. After that, $filterfirst$ applies the conventional $A^*$ algorithm \cite{Dijkstra59} over the remaining road networks (excluding those with obstacles) to find the shortest path between the source, $src$, and the destination, $dst$.
}


\nop{We implement a baseline algorithm, namely $Baseline$, which traverses the graph, starting from $src$, in a breadth-first manner. Below, we will compare our proposed $PRAO$ approach with $Baseline$, in terms of both CPU time and I/O cost. 
}

\begin{table}[t!]
\centering\footnotesize
\begin{tabular}{||p{1.8cm}||p{6 cm}||}
\hline
    {\bf Parameter} & \qquad \qquad\qquad\qquad{\bf Value}\\
     \hline
     \hline
     $\alpha$& 0.1, 0.3,  {\bf0.5}, 0.7, 0.9\\
     \hline
      $\epsilon$ & 20, 30,  {\bf 50}, 70, 90\\
      \hline
      $\sigma$& 5, 8, {\bf10}, 15, 20, 25, 30\\
      \hline      
      $|S|$ & 1, 3, {\bf5}, 8, 10\\
       \hline      
       $d$ & 2, {\bf 5}, 8, 10, 15\\
      \hline
      $|V|$& 10K, 20K, {\bf30K}, 40K, 50K, 200K, 500K, 1M, 1.2M\\
      \hline
\end{tabular}
\caption{Parameter settings}
\label{table:parameter}
\end{table}

\begin{figure}[t!]
\centering
\subfigure[][{\small CPU time}]{                    
\scalebox{0.45}[0.45]{\includegraphics{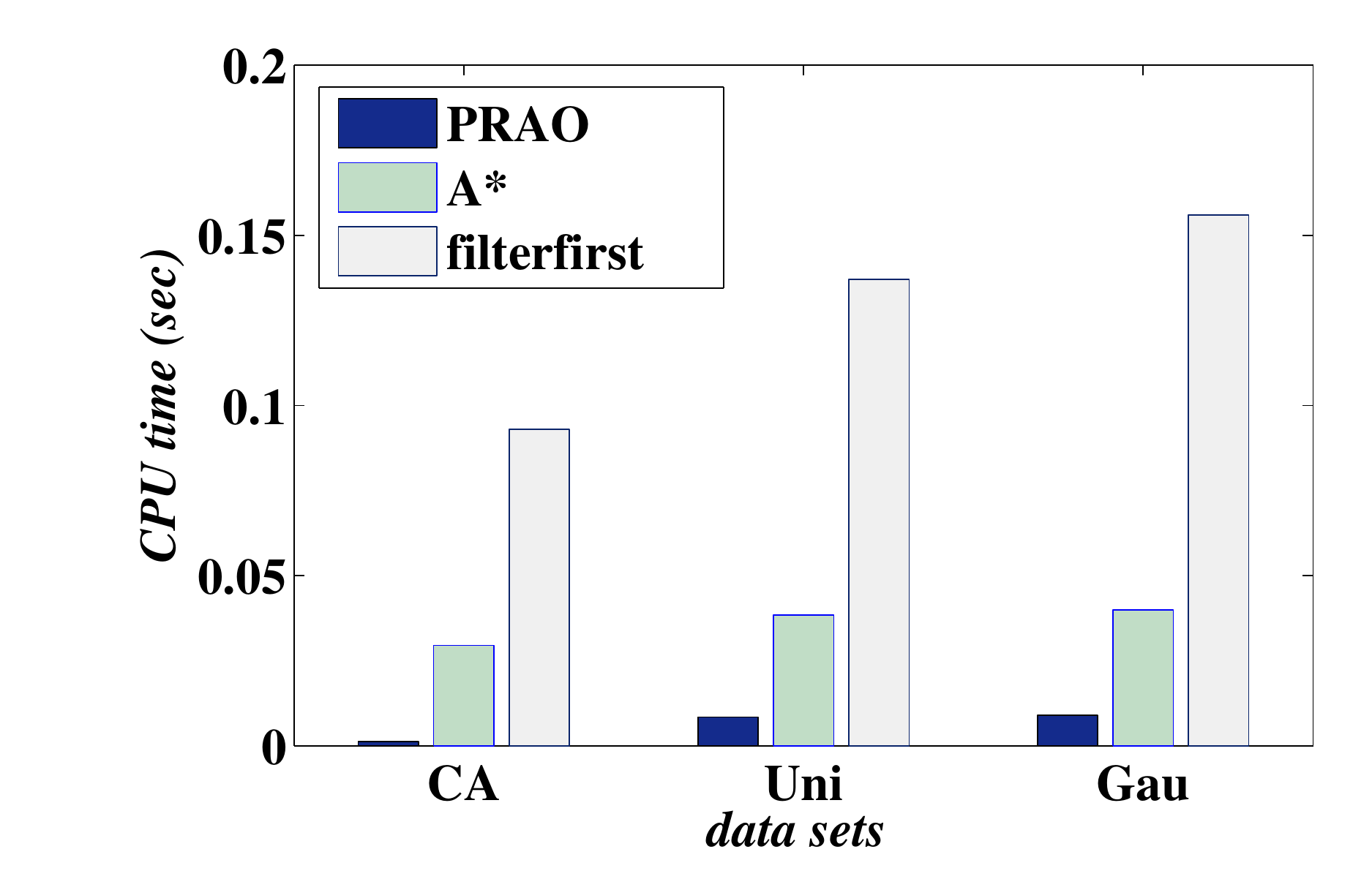}}
}%
\subfigure[][{\small I/O cost}]{                    
\scalebox{0.45}[0.45]{\includegraphics{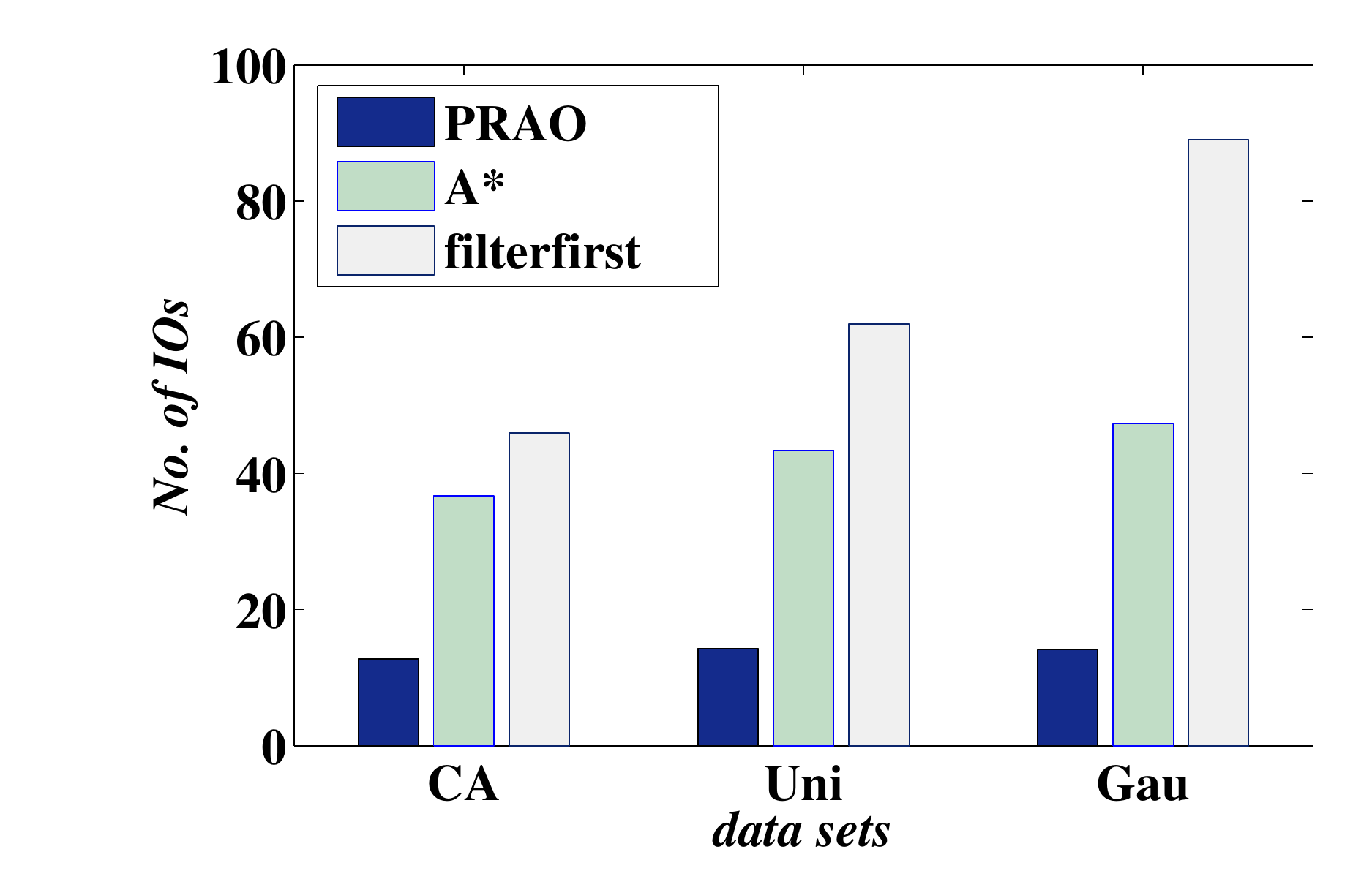}}
}\vspace{-1ex}
\caption{The PRAO efficiency vs. real/synthetic data sets.} \label{exper:PRAOvsBase} \vspace{2ex}
\end{figure}

\vspace{0.5ex}\noindent {\bf Parameter Settings:} Table \ref{table:parameter} depicts the parameter settings in our experiments, where bold numbers are default parameter values. In each set of our subsequent experiments, we will vary one parameter while setting other parameters to their default values. We ran our experiments on a machine with Intel(R) Core(TM) i7-6700 CPU 3.40 GHz (8 CPUs) and 64 GB memory. All algorithms were implemented by C++.

\subsection{The PRAO Performance}
\label{subsec:PRAO_exper_efficiency}


\vspace{0.5ex}\noindent {\bf The PRAO Performance vs. Real/Synthetic Data Sets:} 
{
First, as illustrated in Figure \ref{exper:PRAOvsBase}, we test the performance of our PRAO approach against two baseline algorithms namely $A^*$ and $filterfirst$  with real/synthetic data sets $CA$, $Uni$, and $Gau$.
From figures, we can see that PRAO outperforms both $A^*$ and $filterfirst$ }for all the three data sets, in terms of both CPU time and I/O cost. This is because PRAO applies effective pruning methods with the help of the index. The experimental results confirm the effectiveness of our pruning methods, and efficiency of our PRAO approach.

{
In Figure \ref{exper:PRAOvsBase}, we can see that PRAO outperforms both baselines, $A^*$ and $filterfirst$. Specifically, our PRAO algorithm utilizes a variant of the R$^*$-tree index to facilitate the query processing. On the node level of the index, we can quickly filter out many false alarms of path candidates (which violate the constraints of keyword or weather conditions) with low costs, and obtain a small set of path candidates after the index traversal (i.e., many false alarms are pruned on higher levels with low costs).
In contrast, $A^*$ is based on a variant of the Dijkstra's algorithm starting from the source point to search for paths towards all directions, and many searching paths may not even be towards the destination point, which may consume extra computation costs for searching and refinement.
Similarly, the second baseline, $filterfirst$, takes much computational time in pruning unnecessary edges at the filtering step. Similar to $A^*$, $filterfirst$ struggles from exploring paths that may not be towards the destination.
Therefore, our PRAO algorithm is more efficient than both $A^*$ and $filterfirst$ algorithms, as confirmed by Figure 8.
}

\begin{figure}[t!]
\centering
{                    
\scalebox{0.45}[0.35]{\includegraphics{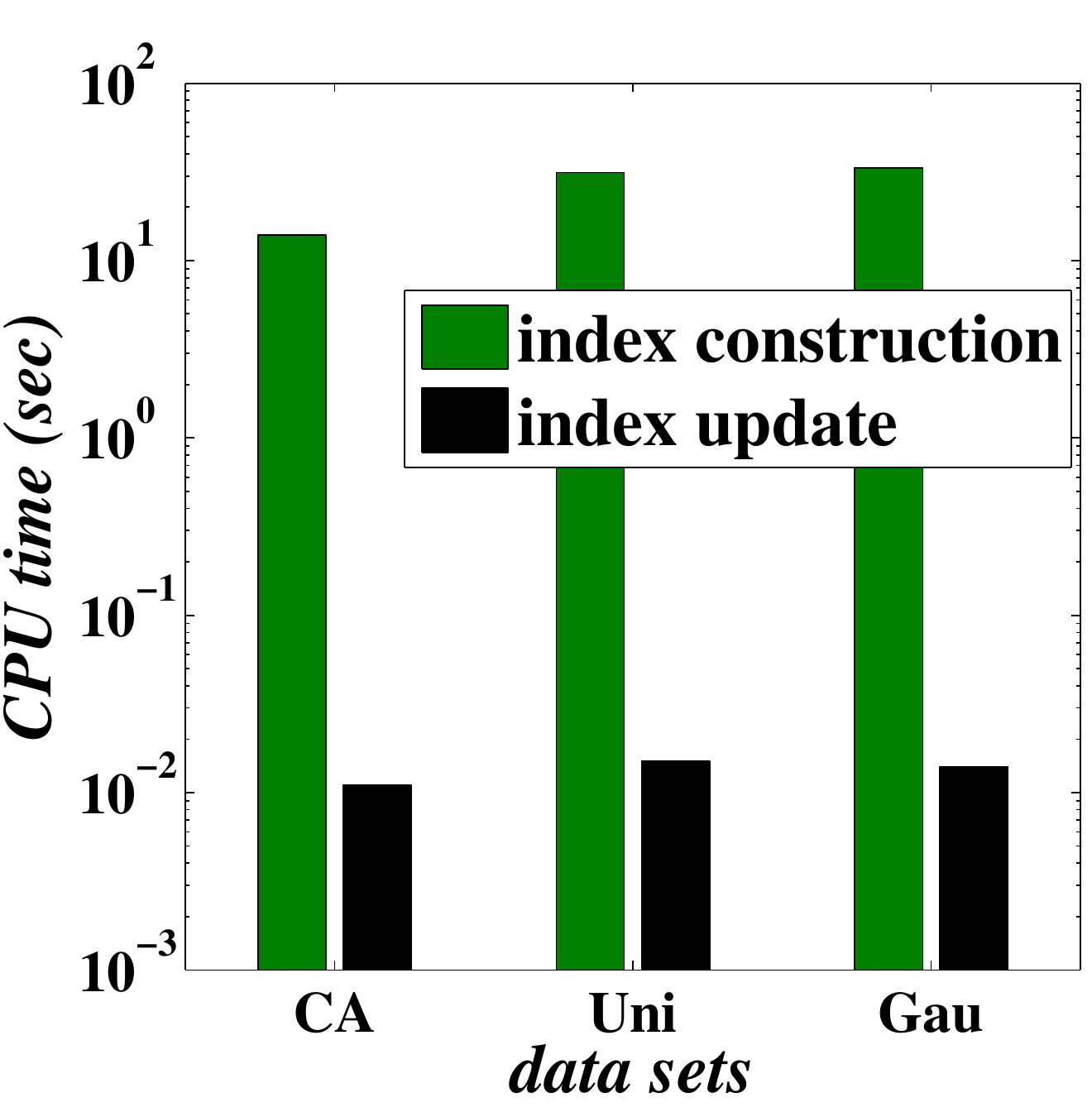}}%
}\vspace{-1ex}
\caption{The index construction and update times vs. real/synthetic data sets.} \label{fig:conVSupdate} 
\end{figure}

\vspace{0.5ex}\noindent {\bf Index Construction and Update Time vs. Real/Synthetic Data Sets:}
Figure \ref{fig:conVSupdate} presents the efficiency of the index construction and dynamic updates (as discussed in Section \ref{subsec:indexing}) over $CA$, $Uni$, and $Gau$ data sets. From experimental results, the index construction takes about 13.9 $sec$ for $CA$ data set, 31.3 $sec$ for $Uni$ data, and 33.2 $sec$ for $Gau$ data. As we discussed earlier, PRAO builds the R$^*$-tree index offline only once. Moreover, we can see that the cost of updating the index is low, for example, 0.011 $sec$ for $CA$ (with $|V|= 21,048$ and $|E|= 21,693$), and about 0.017 $sec$ for $Uni$ and $Gau$ (with $|V|= 30K$ and $|E|= 40K$).



In the sequel, to evaluate the robustness of our PRAO approach, we will vary different parameter values over synthetic data sets and report the experimental results.

\begin{figure}[t!]
  \centering\vspace{-2ex}
    \subfigure[][{\small CPU time}]{                    
    \scalebox{0.45}[0.45]{\includegraphics{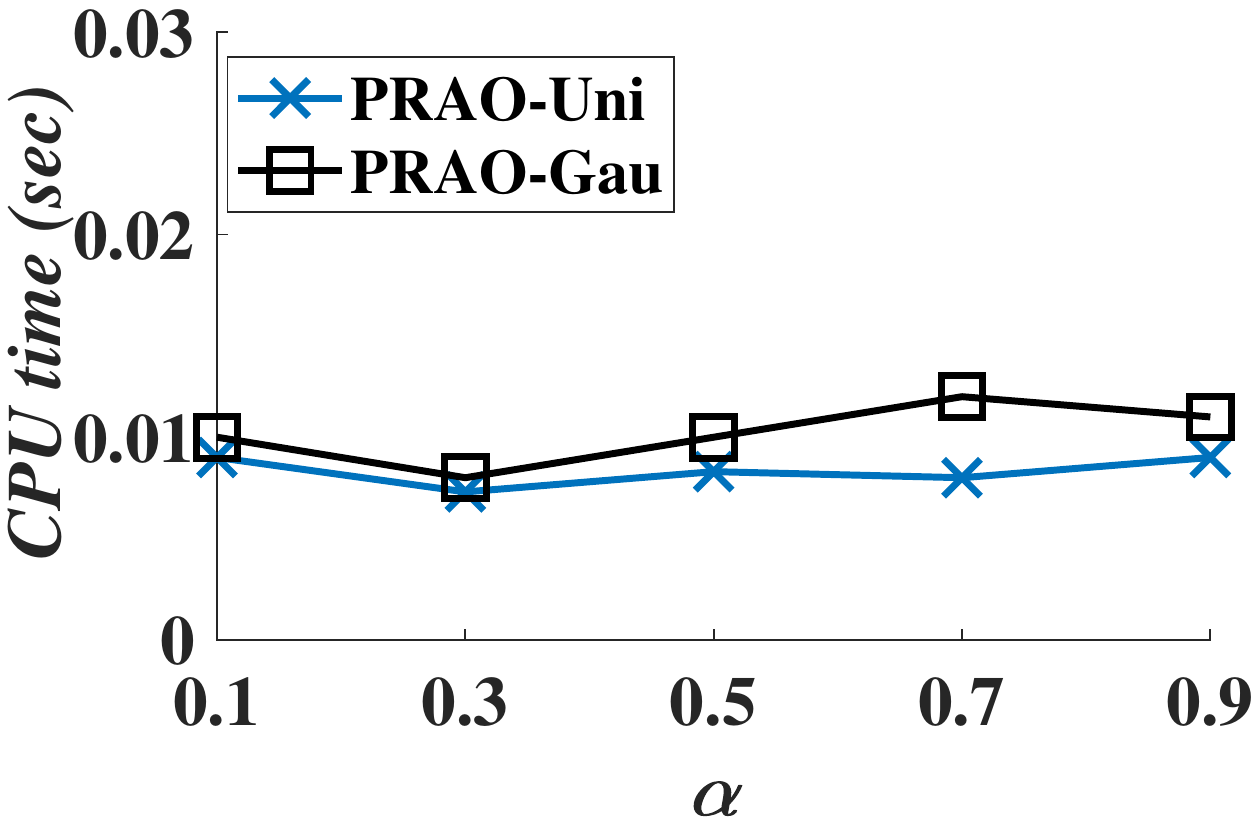}}\label{subfig:alpha_time}
    }%
    \subfigure[][{\small I/O cost}]{                    
    \scalebox{0.45}[0.45]{\includegraphics{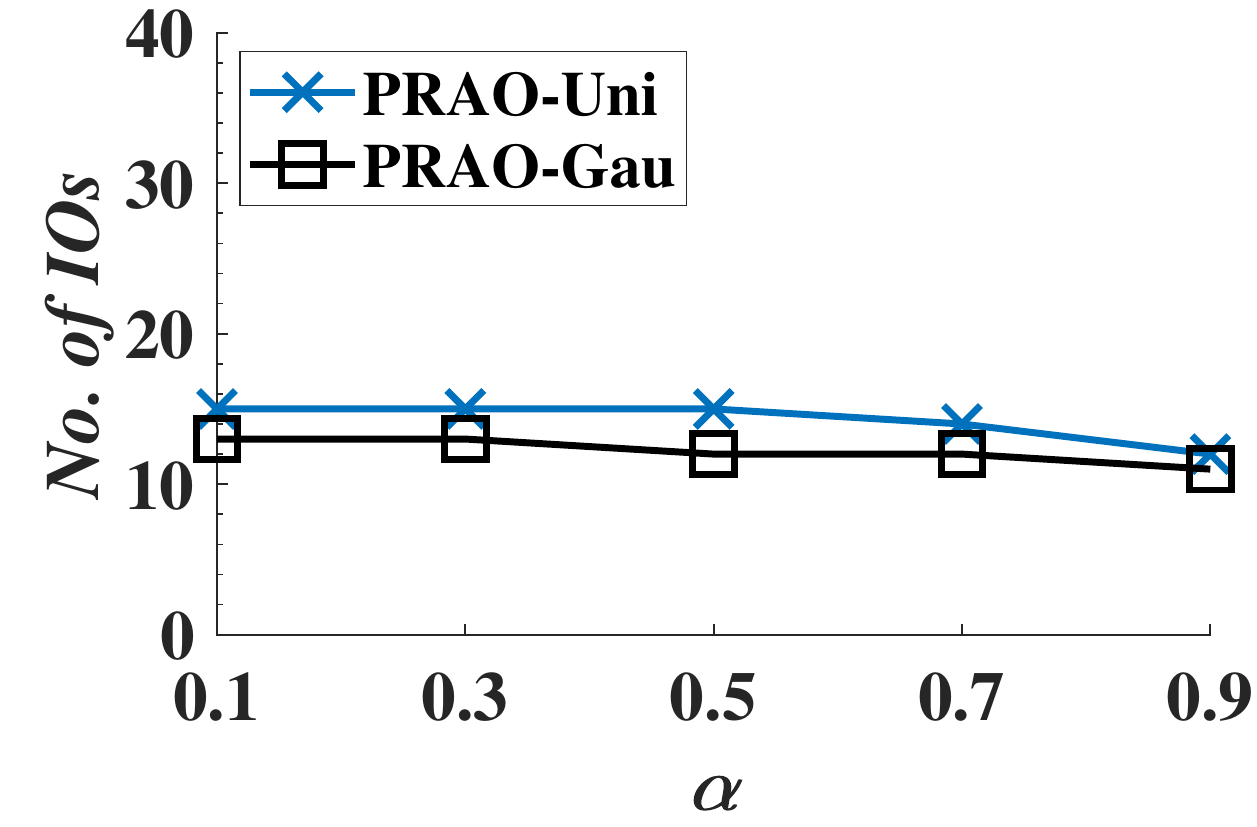}}\label{subfig:alpha_IO}
    }\vspace{-1ex} 
     \caption{The PRAO efficiency vs. probabilistic weather confidence threshold $\alpha$.} 
     \label{fig:alpha}\vspace{2ex}
\end{figure}

\vspace{0.5ex}\noindent {\bf Effect of Probabilistic Weather Confidence Threshold $\alpha$:} Figure \ref{fig:alpha} shows the effect of the probabilistic threshold $\alpha$ for weather conditions on the PRAO performance over $Uni$ and $Gau$ data sets, where $\alpha$ varies from 0.1 to 0.9, and other parameters are set to their default values (see Table \ref{table:parameter}). From figures, for different $\alpha$ values, we can see that the CPU time and I/O cost remain low (e.g., about 0.01 $sec$ and less than 15 I/O accesses). This indicates the efficiency of our PRAO approach against different probabilistic weather threshold $\alpha$.

\begin{figure}[t!]
  \centering\vspace{-2ex}
    \subfigure[][{\small CPU time}]{                    
    \scalebox{0.45}[0.45]{\includegraphics{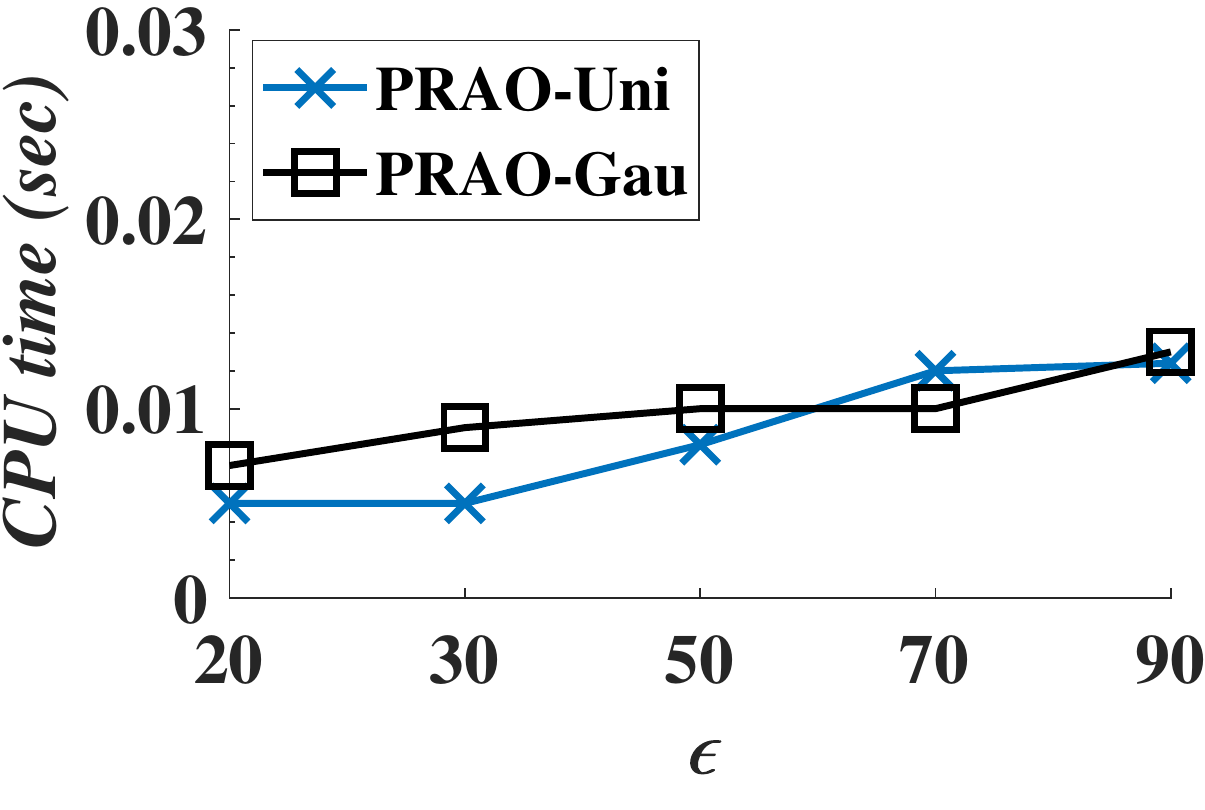}}
    }%
    \subfigure[][{\small I/O cost}]{                    
    \scalebox{0.45}[0.45]{\includegraphics{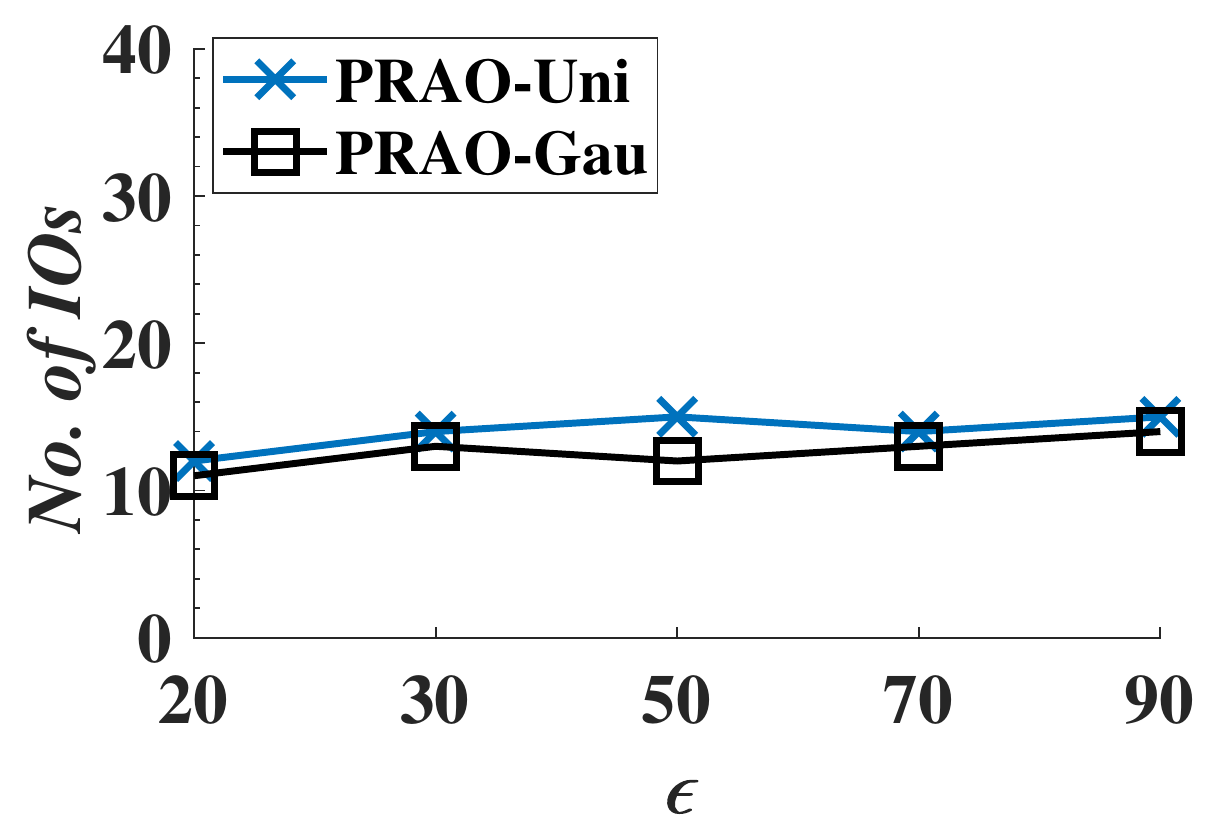}}
    }\vspace{-1ex} 
     \caption{The PRAO efficiency vs. weather obstacle threshold $\epsilon$.} 
     \label{fig:epsilon}\vspace{2ex}
\end{figure}


\vspace{0.5ex}\noindent {\bf Effect of Weather Obstacle Threshold $\epsilon$:} Figure \ref{fig:epsilon} reports the CPU time and I/O cost of our PRAO approach, by varying the weather obstacle threshold $\epsilon$ from 20 to 90. When threshold $\epsilon$ is small, many edges (in turn, paths) will be considered as containing weather-based obstacles, and thus high pruning power can be achieved, which incurs low CPU time and I/O cost. From the experimental results, for different $\epsilon$ values, the PRAO approach is efficient, with low CPU time (i.e., 0.005$\sim$0.015 $sec$) and I/O cost (i.e., 10$\sim$15 I/Os), which confirms the efficiency of our PRAO approach with respect to different $\epsilon$ value.


\balance

\begin{figure}[t!]
  \centering
\subfigure[][{\small CPU time}]{                    
\scalebox{0.4}[0.4]{\includegraphics{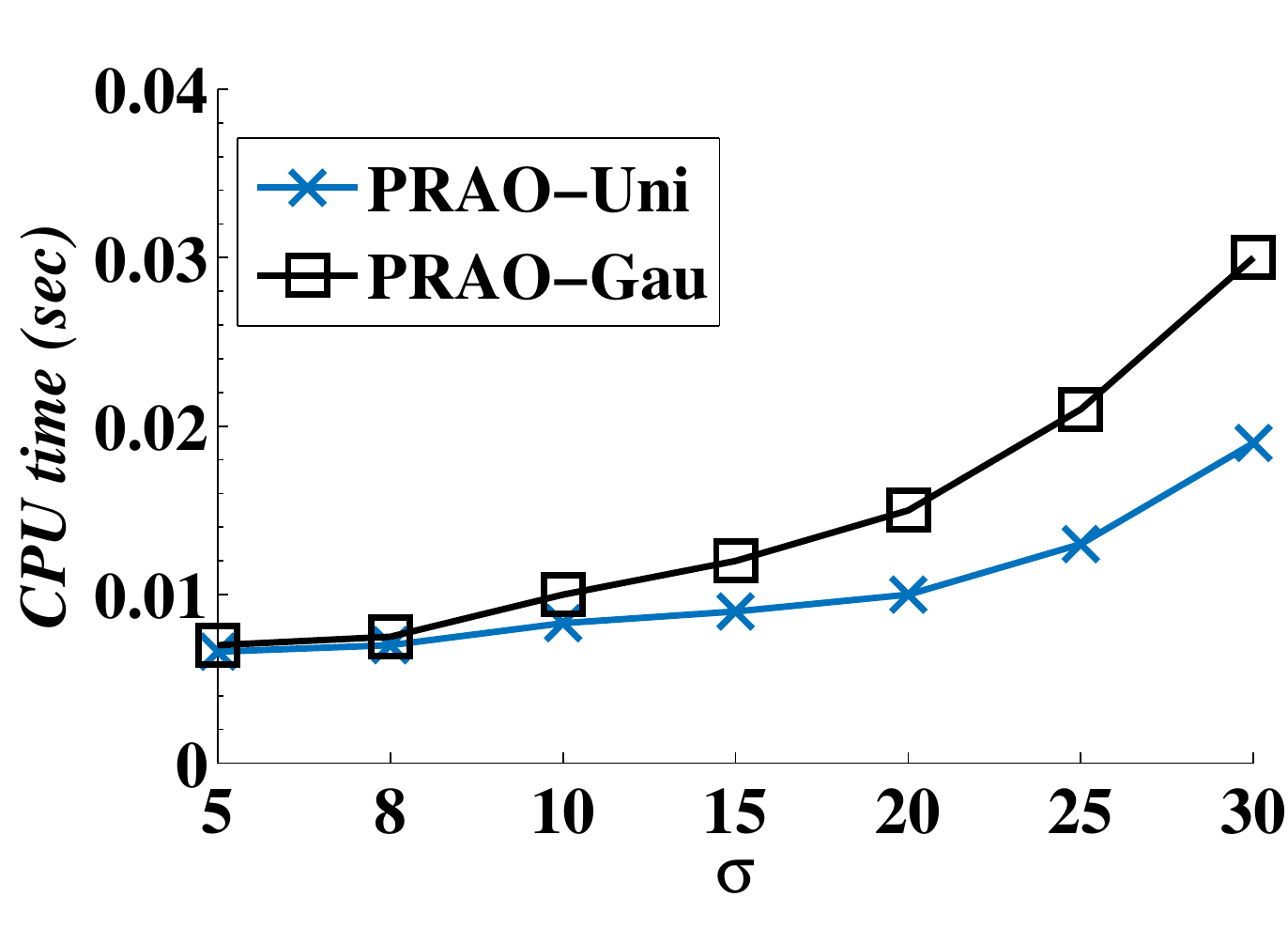}}\label{subfig:sigma_time}
}\quad%
\subfigure[][{\small I/O cost}]{                    
\scalebox{0.42}[0.42]{\includegraphics{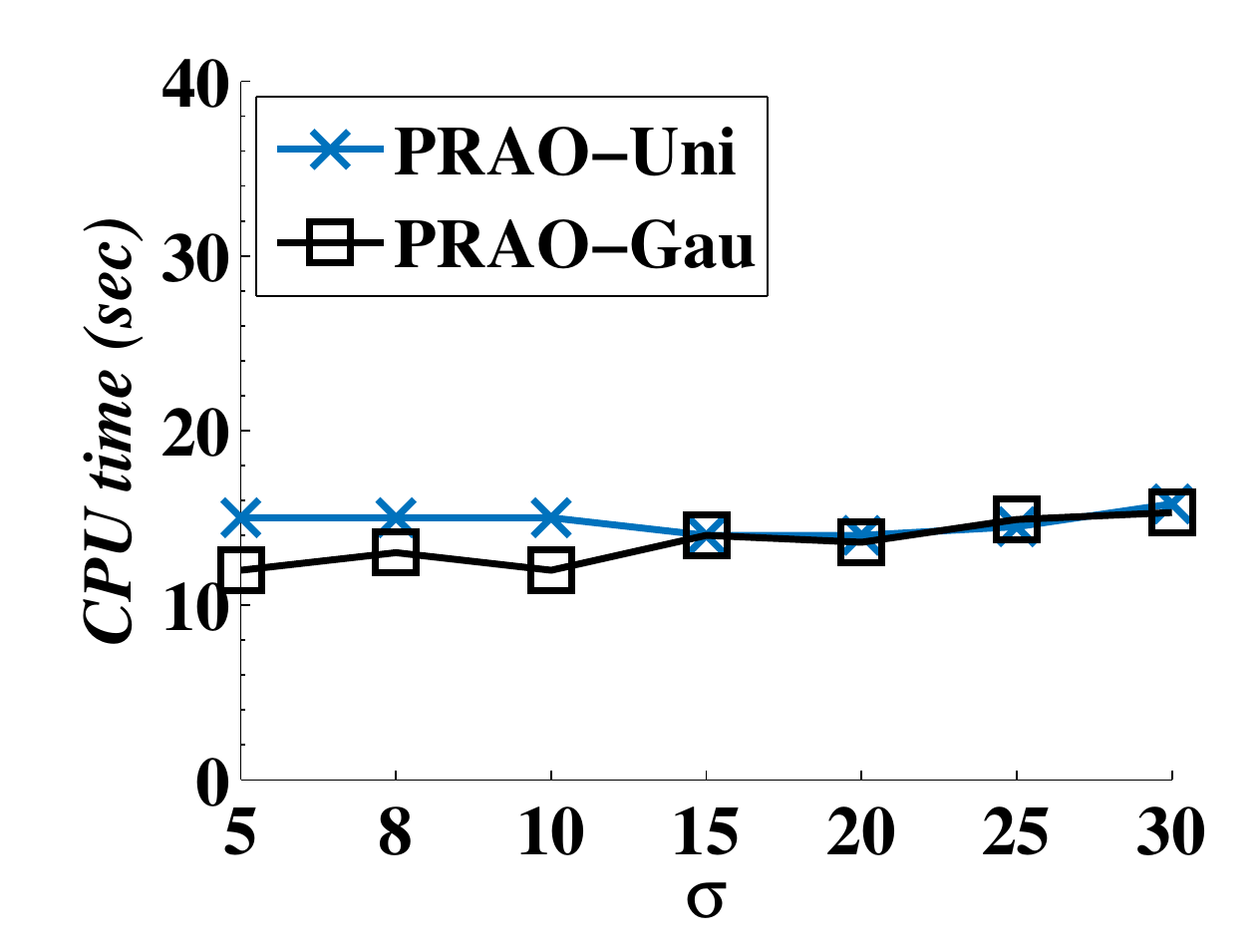}}\label{subfig:sigma_IO}
}\vspace{-1ex}  
     \caption{The PRAO efficiency vs. valid path length $\sigma$.} 
     \label{fig:sigma}\vspace{2ex}
\end{figure}

\vspace{0.5ex}\noindent {\bf Effect of Valid Path Length $\sigma$:} Figure \ref{fig:sigma} varies the average length, $\sigma$, of valid query paths from 5 to 30, where other parameters are set to default values. Intuitively, longer paths from PRAO answer set lead to higher search cost. Thus, when $\sigma$ increases, as shown in Figure \ref{subfig:sigma_time}, the CPU time also becomes higher. On the other hand, the I/O cost is not very sensitive to the $\sigma$ value. From experimental results, the CPU time and I/O cost remain low (i.e., less than 0.03 $sec$ and 10$\sim$15 I/Os, respectively). 

\begin{figure}[t!]
\centering\vspace{-2ex}
\subfigure[][{\small CPU time}]{                    
\scalebox{0.45}[0.45]{\includegraphics{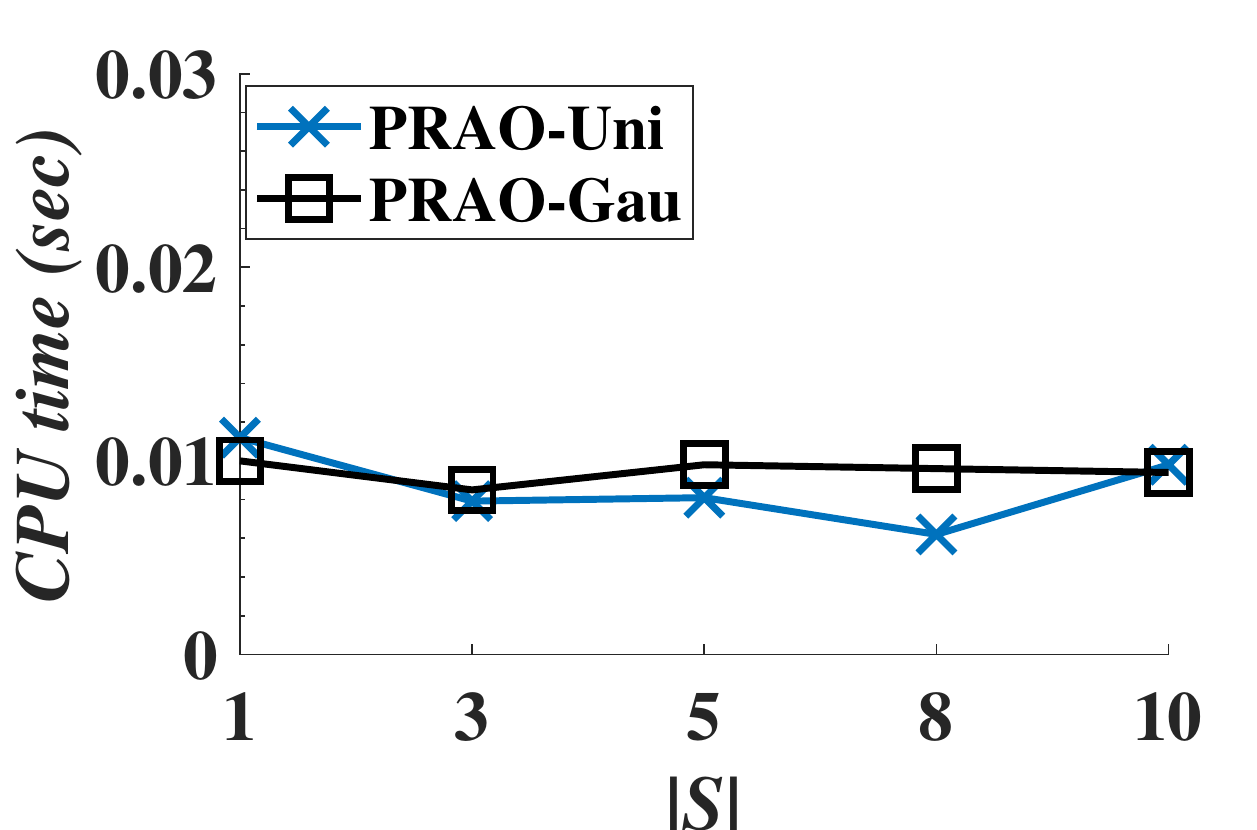}}
}%
\subfigure[][{\small I/O cost}]{                    
\scalebox{0.45}[0.45]{\includegraphics{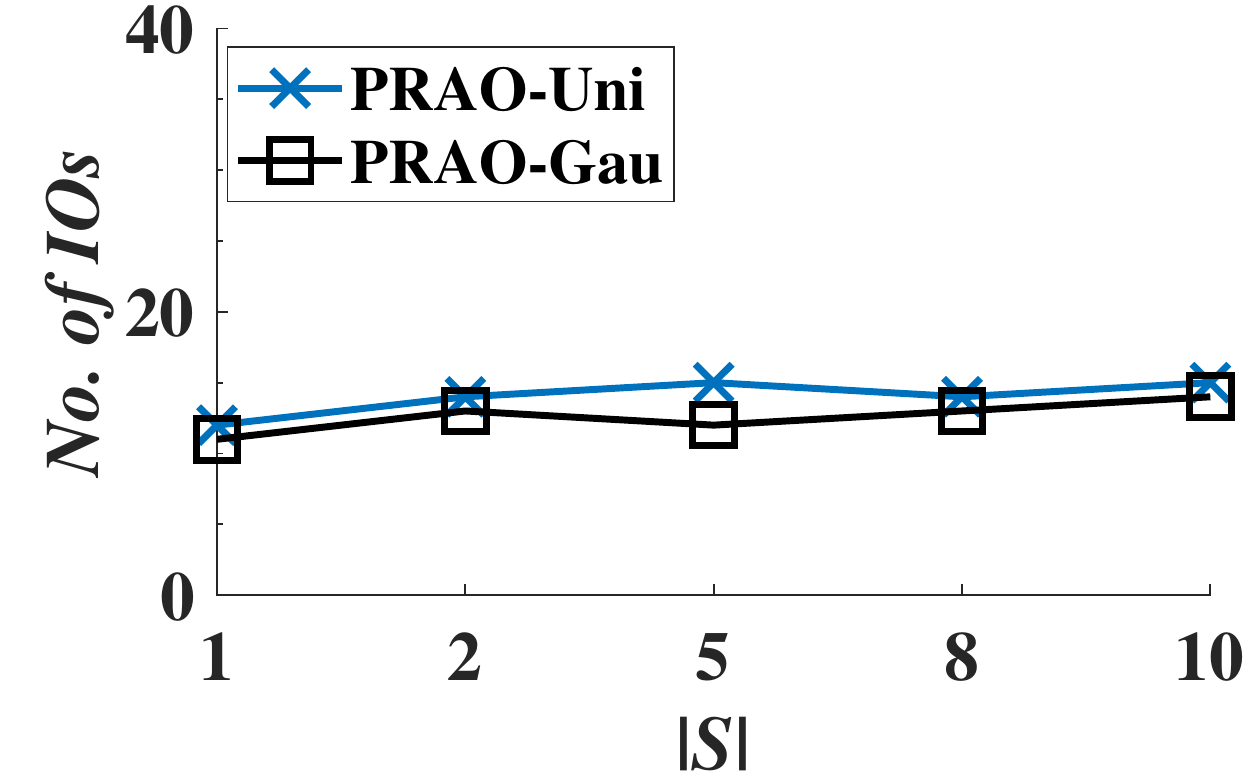}}
}\vspace{-1ex}
\caption{The PRAO efficiency vs. the size, $|S|$, of keyword obstacle set.} \label{fig:|S|}
\end{figure} 

\begin{figure}[t!]
\centering\vspace{-2ex}
\subfigure[][{\small CPU time}]{                    
\scalebox{0.25}[0.34]{\includegraphics{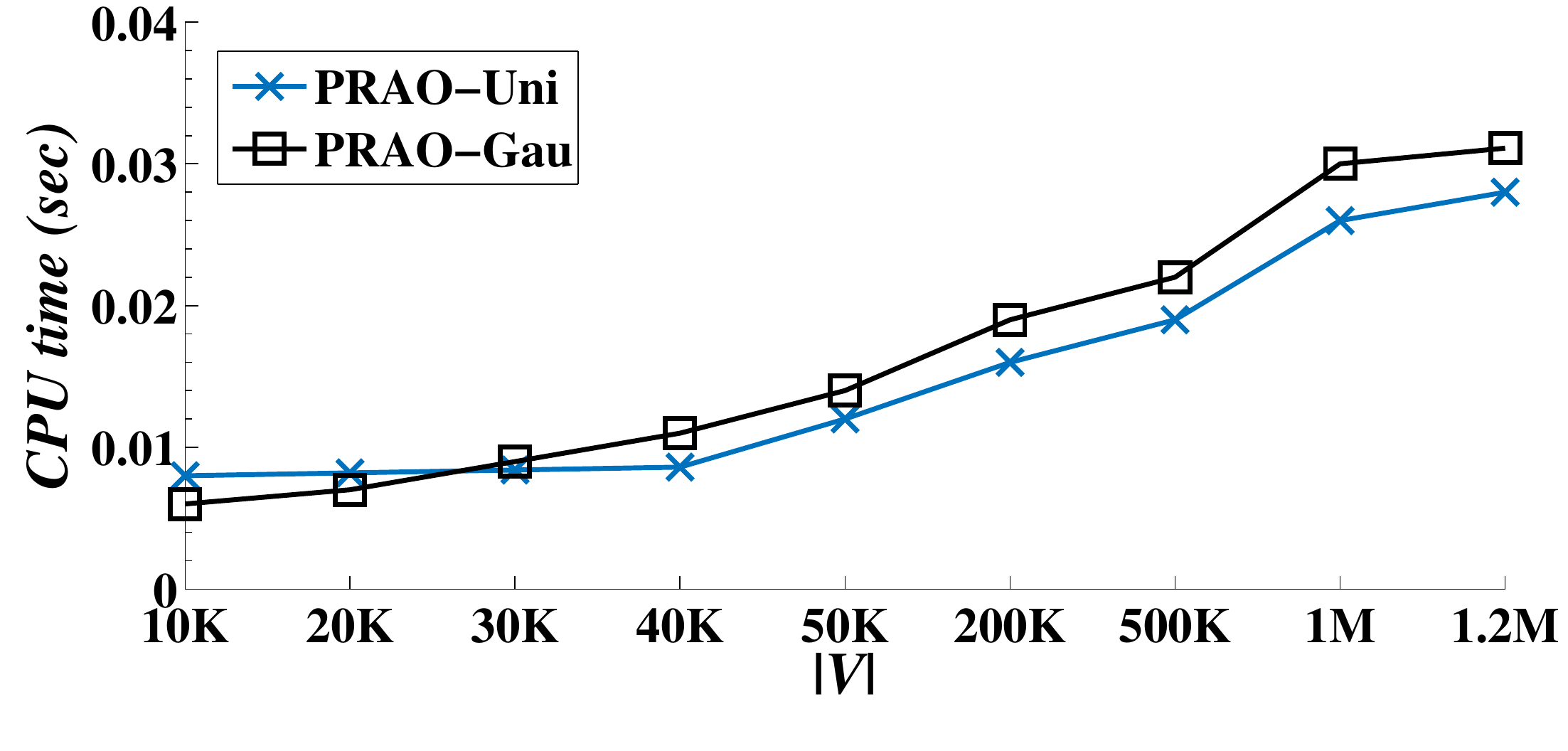}}\label{exper:V}
}%
\subfigure[][{\small I/O cost}]{                    
\scalebox{0.25}[0.34]{\includegraphics{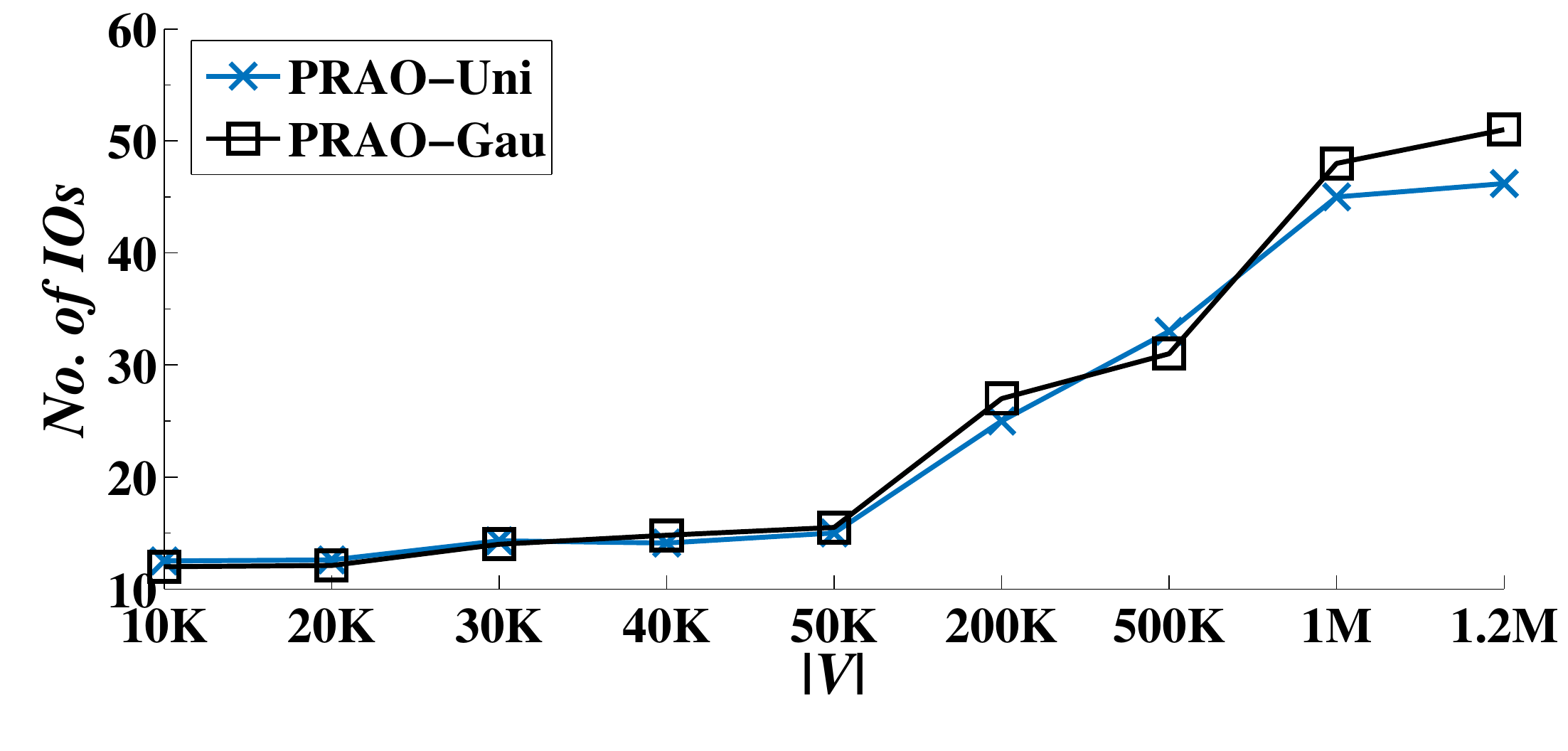}}
}\vspace{-1ex}
\caption{The PRAO efficiency vs. the number, $|V|$, of vertices.} \label{exper:scale} \vspace{2ex}
\end{figure} 

\begin{figure}[t!]
\centering\vspace{-4ex}
\subfigure[][{\small CPU time}]{                    
\scalebox{0.40}[0.40]{\includegraphics{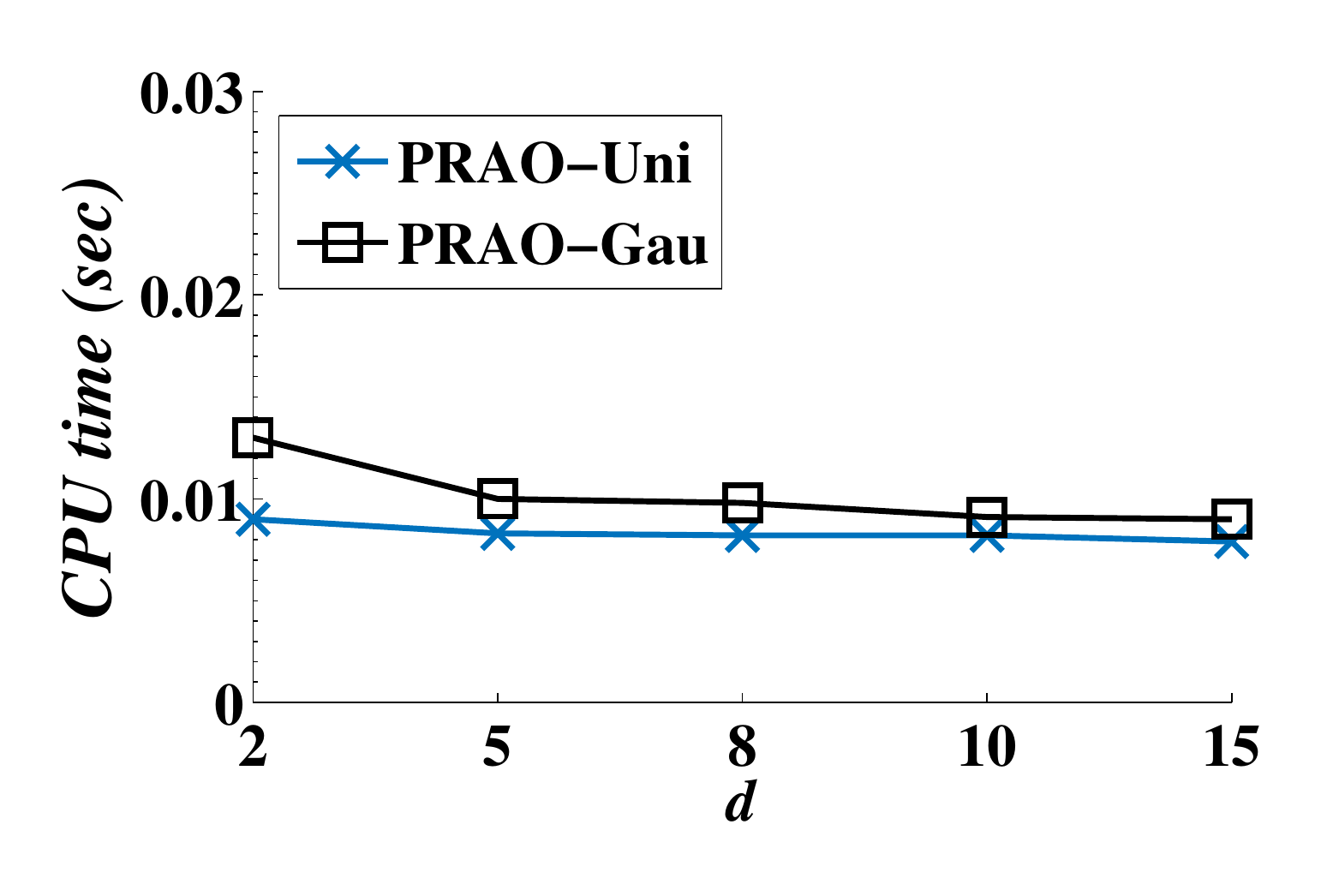}}
}%
\subfigure[][{\small I/O cost}]{                    
\scalebox{0.40}[0.40]{\includegraphics{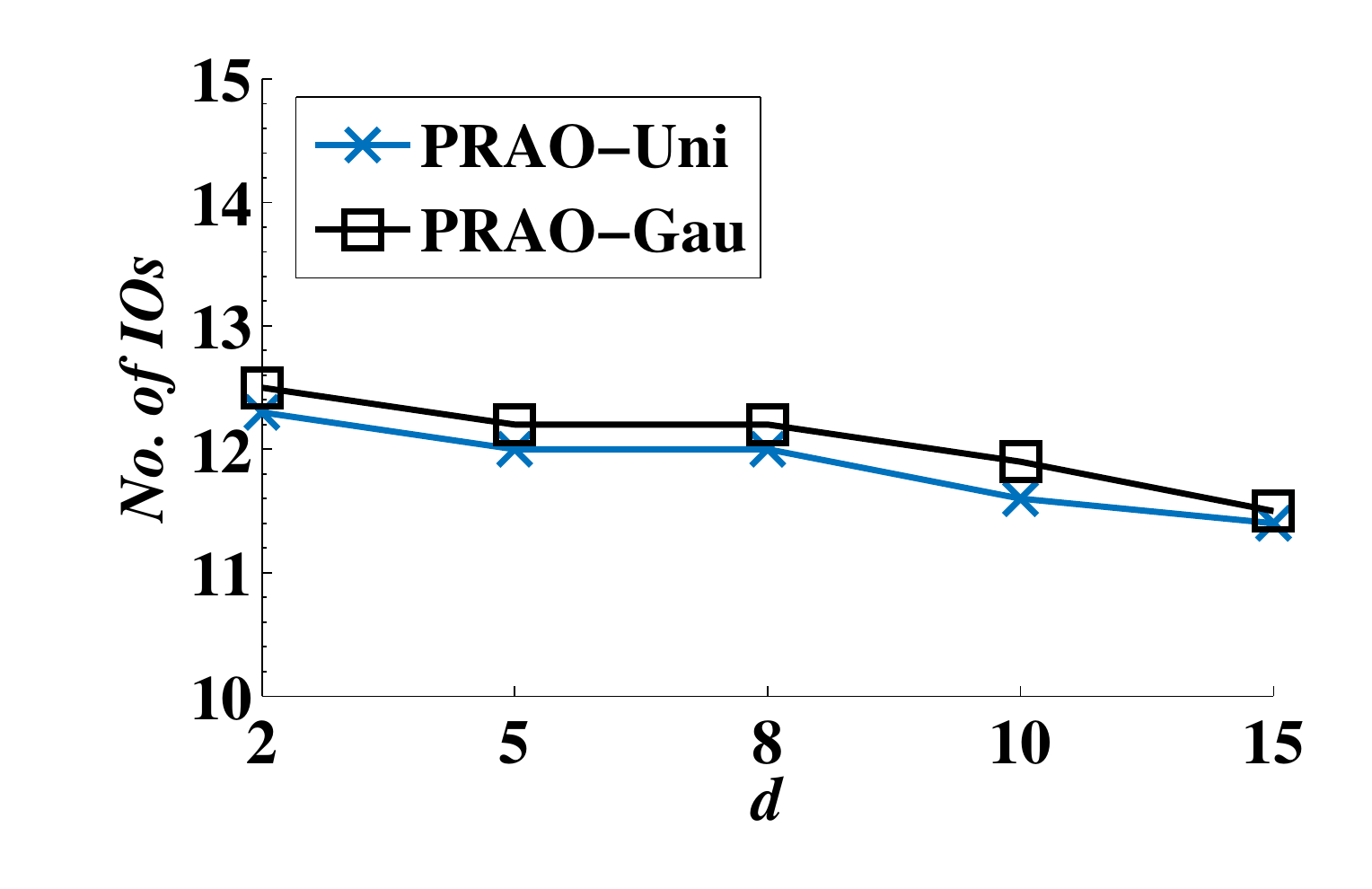}}
}\vspace{-1ex}
\caption{The PRAO efficiency vs. the number, $d$, of pivots.} \label{exper:pivots} 
\end{figure} 

\begin{figure}[t!]
\centering\vspace{-3ex}
\subfigure[][{\small Queue $\mathcal{Q}$}]{                   
\scalebox{0.27}[0.4]{\includegraphics{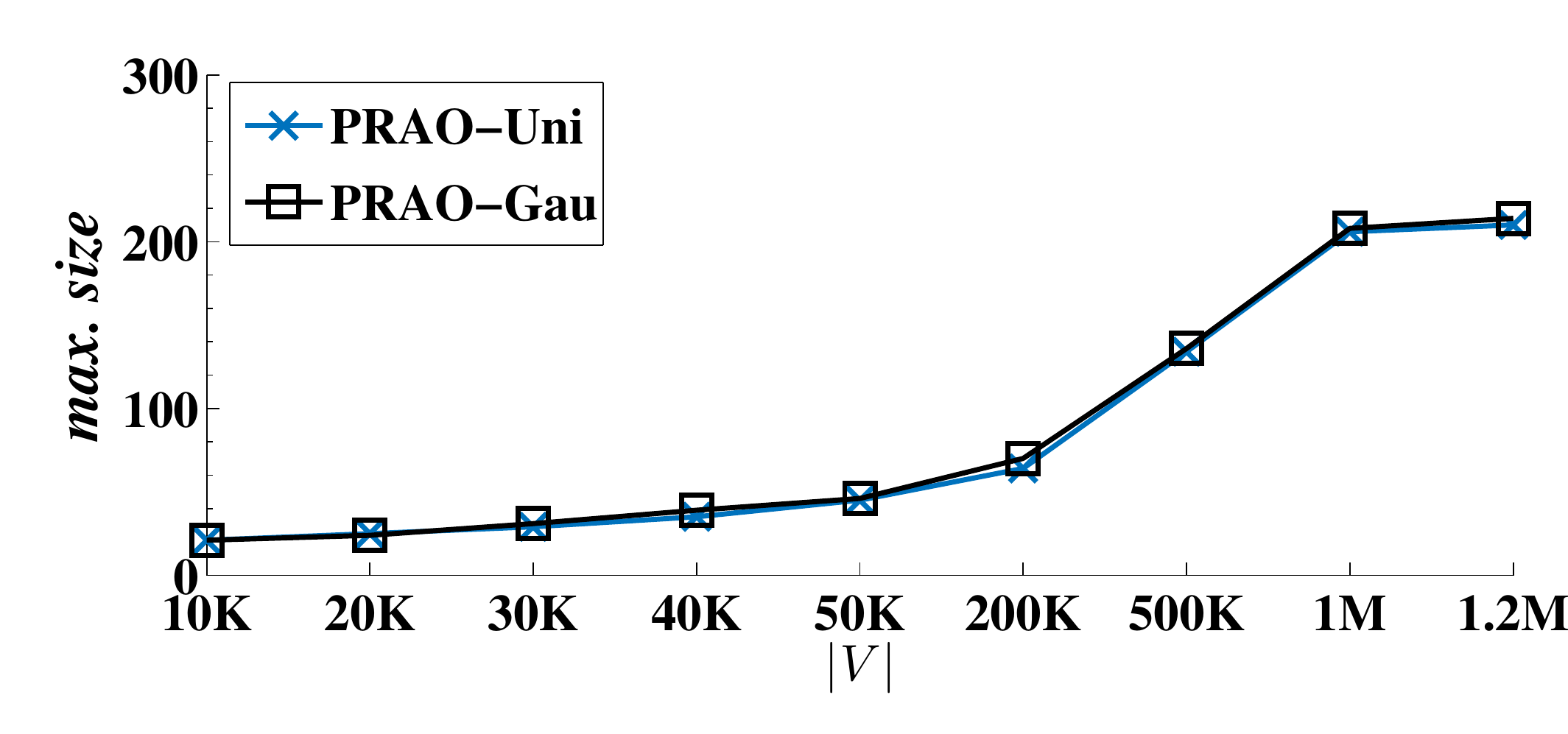}}\label{exper:q_size}
}%
\subfigure[][{\small Heap $\mathcal{H}$}]{                    
\scalebox{0.27}[0.40]{\includegraphics{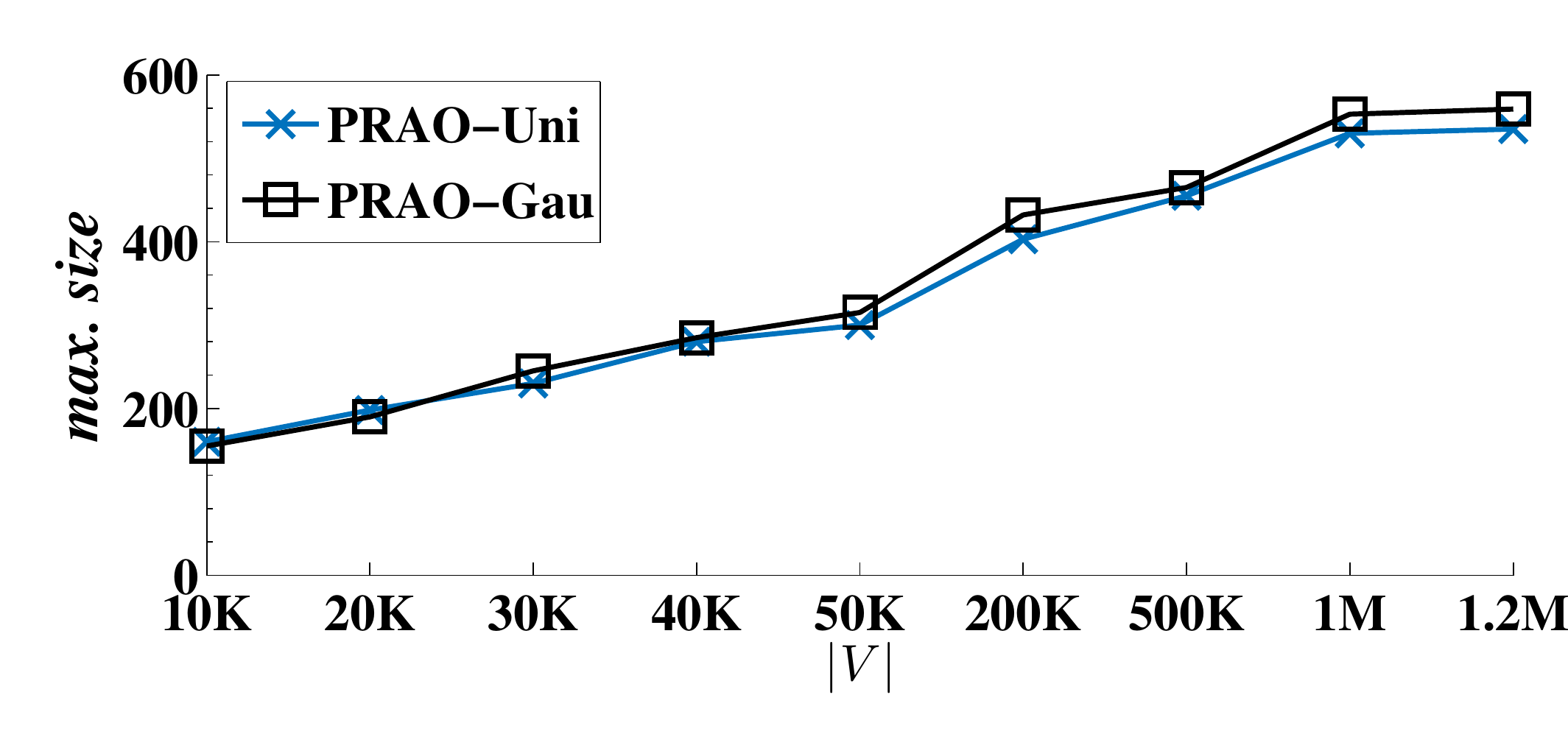}}\label{exper:h_size}
}\vspace{-1ex}
\caption{The maximum sizes of the queue $\mathcal{Q}$ and the heap $\mathcal{H}$ of {\sf PRAO-QP} procedure while processing queries.} \label{exper:h&q_sizes} \vspace{2ex}
\end{figure} 


\vspace{0.5ex}\noindent {\bf Effect of the Size, $|S|$, of Keyword Obstacle Set:} Figure \ref{exper:scale} examines the effect of the number, $|S|$, of ad-hoc obstacle keywords in set $S$, where $|S| = 1, 3, 5, 8,$ and $10$, and default values are used for other parameters. In figures, the CPU time and I/O cost are not very sensitive to the value of $|S|$. The CPU time for different $|S|$ remains low (i.e., around 0.01 $sec$), whereas the number of I/Os is about 10$\sim$15, which indicates the efficiency of our proposed PRAO approach against $|S|$.



{
\vspace{0.5ex}\noindent {\bf Effect of the Number, $|V|$, of Vertices:} Figure \ref{exper:scale} demonstrates the scalability of our PRAO approach, by changing the number, $|V|$, of vertices in road networks $G$ from $10K$ to $1.2M$, where other parameters are set to their default values. With larger road networks (i.e., larger $|V|$), there are more candidate paths between source $src$ and destination $dst$. Therefore, as shown in Figure \ref{exper:V}, the CPU time tends to smoothly increase for larger $|V|$. Both CPU time and I/O cost remain low (i.e., 0.005$\sim$0.033 $sec$ and 11$\sim$51 I/Os, respectively). This confirms the scalability of our proposed PRAO approach against large road networks. 
}

{
\vspace{0.5ex}\noindent {\bf Effect of the Number, $d$, of pivots:} Figure \ref{exper:pivots} illustrates the performance of our PRAO algorithm varying the number, $d$, of pivots from 2 to 15, where other parameters are set to their default values.
From figures, we can see that more pivots lead to better PRAO performance (including CPU time and I/O cost). This is because more pivots will incur tighter bounds for the traveling time pruning (as discussed in Section \ref{subsec:travelingtimepuning}). From experimental results, The CPU time and I/O cost with different numbers of pivots remain low (i.e., below 0.013 $sec$ and below 13 page accesses, respectively).
}

{
\vspace{0.5ex}\noindent {\bf Size Evaluation of Queue $\mathcal{Q}$ and Heap $\mathcal{H}$:}
Next, we study the behaviors of the priority queue $\mathcal{Q}$ and the heap $\mathcal{H}$ during the processing of the PRAO-QP procedure. Figures \ref{exper:q_size} and \ref{exper:h_size} report the maximum sizes of queue $\mathcal{Q}$ and heap $\mathcal{H}$, respectively, during PRAO query processing, where $|V|$ varies from 10$K$ to 1.2$M$, and other parameters are set to their default values. From figures, we can see that, when the size of the data set increases, the maximum sizes of both $\mathcal{Q}$ and $\mathcal{H}$ increase. Nevertheless, even for large data sets, the sizes of $\mathcal{Q}$ and $\mathcal{H}$ remain small. From our experimental results, even with a large road network with 1.2$M$ vertices, the maximum sizes of $\mathcal{Q}$ and $\mathcal{H}$ are less than 211 and 600, respectively. This confirms the space efficiency of our PRAO query algorithm that utilizes data structures $\mathcal{Q}$ and $\mathcal{H}$ to enable efficient PRAO query processing. 
}

We also tested other parameters (e.g., the number, $|E|$, of edges) and data distributions (e.g., skewed vertex, or obstacle keyword distributions), and do not report similar experimental results. In summary, our PRAO approach can achieve good performance, in terms of both CPU time and I/O cost.



\section{Related Work}
\label{sec:related}

\vspace{0.5ex}\noindent {\bf Query Processing on Road Networks:} There are many existing works on the data model of spatial road networks
(or graphs), whose road segments (edges) are associated with deterministic weights. The \textit{shortest path query} is a classical query over road
networks, which retrieves the shortest path between source and destination on road networks. Dijkstra \cite{Dijkstra59} proposed the well-known
Dijkstra's algorithm to solve such a problem. In order to improve the query efficiency, several variants have been proposed to heuristically prune the search space \cite{Kung86} or materialize some paths \cite{agrawal1990direct,Ioannidis93, goldberg2005computing}. Huang et al. \cite{huang1997integrated} studied the shortest path search, by verifying spatial constraints (e.g., altitude) of the passing areas through the join operator. Li et al. \cite{Li05b} explored the shortest path queries between source and destination that pass through some types of interesting data points.
Song et al. \cite{A1} studied the repairing of inconsistent timestamps that do not conform to the required temporal constraints.
Neighborhood constraints have been studied in \cite{B1, B2}, where the conflict in roads is detected.
Wang et al. \cite{C1, C2}, considered constraints with AND/XOR semantics, in addition to the pairwise constraints. This work filled the missing events referring to the network constraints.
Another graph repair approach was proposed in \cite{D1} that considers not only the constraints in network structure, but also the names (labeling) of events.
Different from prior works, this paper studies obstacle-based shortest path search problem (i.e., PRAO), by considering both ad-hoc keyword-based and weather-based obstacles in the future during the path search, which have not been studied before. Thus, we cannot borrow previous techniques to answer PRAO queries.

Apart from the shortest path query, many other query types have been studied in spatial road networks, for example, range queries \cite{Papadias03b,jeung2010path}, $k$-nearest neighbor ($k$NN) queries\cite{shahabi2003road,Papadias03b}, $k$-nearest neighbor ($k$NN) queries with the incorporation of social influence \cite{R6},
aggregate nearest neighbor query \cite{yiu2005aggregate, papadias2005aggregate}, reverse nearest neighbor queries \cite{Yiu05}, multi-source skyline queries \cite{deng2007multi}, and so on. 
{ Furthermore, the similarity search over uncertain/certain graph databases has been extensively studied in \cite{R1, R2, R3, R5}, where the goal of this query is to find a set of subgraphs from the graph database that are similar to the subgraph query. Prior works on uncertain graphs follow \textit{filtering-and-verification} framework. In the filtering step, different pruning techniques have been proposed to reduce the search space. In particular, they usually designed tight upper/lower bounds of the subgraph similarity that can be utilized to prune unmatched graphs. The verification step is used to compute the actual similarity between a candidate probabilistic graph $g$ and the query graph $q$. However, pruning methods proposed in \cite{R1, R3, R5} cannot be applied to PRAO problem, since our PRAO query is a path routing problem (rather than a probabilistic (sub)graph matching problem). Thus, the pruning methods that prune (sub)graphs (not matching with a query graph) in \cite{R1, R3, R5} cannot be directly used for pruning a path from a source to a destination.
Similarly, \cite{R2} retrieves the matching subgraphs over certain graph database. In this work, authors aim to scale the subgraph matching problem to very large datasets using the parallelism.  The proposed techniques are for certain datasets under the distributed settings, and cannot be applied to solve our PRAO problem over road networks with ad-hoc probabilistic weather information.

There is another query type, keyword search, over uncertain/certain graph databases, which retrieves a set of nodes with certain keywords from the graph. Yuan et al. \cite{R4} proposed solutions to the keyword search following the \textit{filtering-and-verification} framework.
The filtering process uses three pruning techniques, \textit{existence probability pruning, path-based probability pruning, and tree-based probability pruning}. In fact, these pruning techniques cannot be applied to our PRAO problem, since they do not consider ad-hoc weather constraints, and the returned answers are subgraphs (instead of a path) containing keywords (rather than excluding obstacle keywords).
Moreover, in PRAO, the value of weather condition of an edge changes over time, for example, an edge might not satisfy weather condition at 9:00am, but it may be safe to be taken at 9:15am. Similarly, Yuan et al. \cite{R7} considered certain data sets over the distributed environment,
whereas our PRAO problem is considered over a centralized machine and it involves the probabilistic weather condition (e.g., the temperature is 50$^{\circ}$ at 8am with probability 0.6)."
}

In these works, road networks are usually modeled by weighted graphs, and the traveling distance/time is often used to measure the path length between two points on road networks. In addition to weights on edges, in this paper, we also model real scenarios of road networks, by including the properties of road segments (i.e., keywords) and the predicted weather conditions on road networks, which integrate many heterogeneous road-network data. Therefore, with different query types and data models for road networks, we cannot directly apply previous solutions to solve our PRAO problem.


\vspace{0.5ex}\noindent {\bf Probabilistic Graph Management:} In the literature, there are some query types, such as path queries \cite{Nie08,hua2010probabilistic}, reachability queries \cite{Jin11}, and trip planning \cite{Lian14}, over road networks, by considering uncertain traffic conditions or the availability of road segments (e.g., roads under construction). In these works, road networks are modeled by probabilistic graphs, where either edges are associated with uncertain velocity samples of vehicles, or edges have existence probabilities (i.e., edges may exist or not exist). In contrast, in this paper, we consider uncertain weather conditions at vertices of road networks (e.g., wind speed 10mph with the forecasting accuracy or probability 0.8), and utilize the known weather values at vertices to estimate (unknown) weather values at some points on edges (roads), associated with confidences. Moreover, our PRAO query predicates include avoiding ad-hoc weather-based obstacles on the path, which takes into account uncertain weather conditions, and has not been studied before. Our work thus needs to design novel weather-based pruning, specific for PRAO, to enable efficient query answering. 

\vspace{0.5ex}\noindent {\bf Queries in the Presence of Obstacles:} There are some works  \cite{Lozano-PerezW79,Zhang04,Nutanong10,Gao09} on queries taking into account physical obstacles (e.g., buildings and lakes) in the Euclidean space, so that the traveling paths should be detoured. For example, Zhang et al. \cite{Zhang04} studied queries such as the range search, nearest neighbors, $e$-distance joins, and closest pairs in the Euclidean space and with physical obstacles. Due to such obstacles, visible nearest neighbor queries \cite{Nutanong10} or visible reverse nearest neighbor queries \cite{Gao09} are proposed, which retrieve nearest neighbors (NNs) or reverse nearest neighbors (RNNs) of a query point $\mathcal{Q}$ that are visible to $q$ (i.e., without obstacles between $q$ and NNs/RNNs). However, existing works usually considered static physical obstacles and/or obstacle-aware queries in the Euclidean space. In contrast, in our PRAO problem, users can arbitrarily specify \textit{ad-hoc} obstacles on roads (including both keyword and weather obstacles) rather than \textit{static} ones, and furthermore our PRAO queries are performed in graphs (road networks) to retrieve obstacle-aware shortest paths, instead of the Euclidean space. Therefore, we cannot directly use previous methods to tackle our PRAO problem.

\vspace{-2ex}

\section{Conclusion}
\label{sec:conclusion}
In this paper, we formalize and tackle the problem of \textit{path routing over road networks with ad-hoc obstacles} (PRAO), which retrieves paths from source to destination with the smallest traveling time, by avoiding ad-hocly specified keyword-based and weather-based obstacles. To efficiently process the PRAO query, in this paper, we design effective pruning methods (w.r.t. keywords, weather conditions, and traveling times) to filter out false alarms, and propose efficient indexing and query processing algorithms to answer PRAO queries. Extensive experiments have demonstrated the efficiency and effectiveness of our proposed PRAO approach over both real and synthetic data sets.

\vspace{-2ex}
\section*{Acknowledgement}
\label{sec:acknowledgement}
Xiang Lian is supported by NSF OAC No. 1739491 and Lian Startup No. 220981, Kent State University. En Cheng is supported by UA Startup No. 207993, The University of Akron.

\vspace{-2ex}

\balance

\let\xxx=\bibitem\def\bibitem{\par\vspace{-0.4mm}\xxx}
{
\bibliography{ACM_proc}
}

\newpage
\section*{Appendix}
{\small
\subsection*{A. Proof of Lemma \ref{lemma:lem1} }

\noindent {\bf Proof.} Since $e_i.K \bigcap S \neq \emptyset$ holds, we can infer that edge $e_i$ is associated with some obstacle keywords in $S$. Therefore, edge $e_i$ cannot appear on any returned path answer from $src$ to $dst$, based on the PRAO problem definition (given in Definition \ref{def6}). Hence, edge $e_i$ can be safely pruned. \quad $\square$\\

\subsection*{B. Proof of Lemma \ref{lemma:lem2}}
\noindent {\bf Proof.} From the lemma assumption, we have
$UB\_Pr\{W\_val(o_l,t_l)\leq\epsilon\}< 1-\alpha$. Then , we can derive: 
\begin{eqnarray}
&&UB\_Pr\{W\_val(o_l, t_l)\leq\epsilon\}< 1-\alpha \nonumber\\  
&\Longrightarrow& Pr\{W\_val(o_l, t_l)\leq\epsilon\} \leq UB\_Pr\{W\_val(o_l, t_l)\leq\epsilon\}< 1-\alpha\nonumber\\
&\Longleftrightarrow& 1- Pr\{W\_val(o_l, t_l)\leq\epsilon\} > \alpha \nonumber\\
&\Longleftrightarrow& Pr\{W\_val(o_l, t_l)>\epsilon\} > \alpha.\nonumber
\end{eqnarray}

\nop{\begin{eqnarray}
&&UB\_Pr\{W\_val(o_l, t_l)\leq\epsilon\}+UB\_Pr\{W\_val(o_l, t_l)>\epsilon\}\geq 1 \nonumber\\  
&\Longleftrightarrow& 1\leq UB\_Pr\{W\_val(o_l, t_l)\leq\epsilon\}+UB\_Pr\{W\_val(o_l, t_l)\nonumber\\ 
&& >\epsilon\}<UB\_Pr\{W\_val(o_l, t_l)>\epsilon\}+1-\alpha\nonumber\\
&\Longleftrightarrow& UB\_Pr\{W\_val(o_l, t_l)>\epsilon\} > \alpha.\nonumber
\end{eqnarray}}

According to Definition \ref{def:weather_obstacle}, edge $e_i$ contains a weather-based obstacle $o_l$ at timestamp $t_l$, and thus can be safely pruned. \qquad $\square$\\

\subsection*{C. Proof of Lemma \ref{lemma:lem3}}
\noindent {\bf Proof.} Let $UB\_T({\it best\text{-}path\text{-}so\text{-}far})=(e_1.w+e_2.w+\ldots+e_{|{\it best\text{-}path\text{-}so\text{-}far}|}.w)$, where $e_1$ $(=src\rightarrow v_j)$ and $e_{|{\it best\text{-}path\text{-}so\text{-}far}|}$ $(=v_k\rightarrow dst)$. Suppose that, the lower bound traveling time of the path, $Path$, is $LB\_T(Path)=(e_1.w+e_2.w+\ldots+e_c.w)$, where $e_1$ $(=src\rightarrow v_i)$ and $e_c$ $(=v_f\rightarrow dst)$. Since $UB\_T$({\it best-path-so-far}) $<$ $LB\_T(Path)$ holds, by the inequality transitivity, we have $T$({\it best-path-so-far}) $\leq$ $UB\_T$({\it best-path-so-far}) $<$ $LB\_T(Path) \leq T(Path)$. Therefore, the traveling time of the best-so-far path is less than that of $Path$. Thus, $Path$ cannot be the one with the smallest traveling time, and can be safely pruned. \qquad $\square$
 
}


\end{document}